\documentclass[journal]{IEEEtran}

\usepackage{array}
\usepackage{graphicx}
\usepackage{balance} 
\usepackage{url}
\usepackage{float}
\usepackage{amsmath,graphicx,amsfonts,amssymb,url}
\usepackage{setspace}
\usepackage[caption=false]{subfig}
\usepackage{blindtext}
\usepackage[inline]{enumitem}
\usepackage{color}
 \usepackage{subfig}
\usepackage{multirow}
 
\newsavebox{\measurebox}
\usepackage{svg}

\ifCLASSINFOpdf
\else
\fi
\begin{document}
%

\title{Quality Assessment of Free-viewpoint Videos by Quantifying the Elastic Changes of Multi-Scale Motion Trajectories}
 
\author{Suiyi~Ling,~\IEEEmembership{Student Member,~IEEE,}
Jing~Li, Zhaohui Che, Xiongkuo Min,
\\ Guangtao Zhai,~\IEEEmembership{Member,~IEEE,} and~Patrick~Le Callet,~\IEEEmembership{Fellow,~IEEE}

\thanks{Suiyi Ling and Jing Li make equal contributions. Jing Li is the corresponding author.}
\thanks{Suiyi Ling, Jing Li, and Patrick Le Callet are with the {\'E}quipe Image, Perception et Interaction, Laboratoire
des Sciences du Num{\'e}rique de Nantes, Universit{\'e} de Nantes, France (e-mail: suiyi.ling@univ-nantes.fr; jing.li.univ@gmail.com; patrick.lecallet@univ-nantes.fr).}
\thanks{Zhaohui Che, Xiongkuo Min, and Guangtao Zhai are with the Department of Electronic Engineering, Shanghai Jiao Tong University, China (e-mail: \{chezhaohui, minxiongkuo, zhaiguangta\}@sjtu.edu.cn) }
\thanks{Manuscript submitted March, 2019.}

}

\maketitle

\begin{abstract}
Virtual viewpoints synthesis is an essential process for many immersive applications including Free-viewpoint TV (FTV). A widely used technique for viewpoints synthesis is Depth-Image-Based-Rendering (DIBR) technique. However, such technique may introduce challenging non-uniform spatial-temporal structure-related distortions. Most of the existing state-of-the-art quality metrics fail to handle these distortions, especially the temporal structure inconsistencies observed during the switch of different viewpoints. To tackle this problem, an elastic metric and multi-scale trajectory based video quality metric (EM-VQM) is proposed in this paper. Dense motion trajectory is first used as a proxy for selecting temporal sensitive regions, where local geometric distortions might significantly diminish the perceived quality. Afterwards, the amount of temporal structure inconsistencies and unsmooth viewpoints transitions are quantified by calculating 1) the amount of motion trajectory deformations with elastic metric and, 2) the spatial-temporal structural dissimilarity. According to the comprehensive experimental results on two FTV video datasets, the proposed metric outperforms the state-of-the-art metrics designed for free-viewpoint videos significantly and achieves a gain of 12.86\% and 16.75\% in terms of median Pearson linear correlation coefficient values on the two datasets compared to the best one, respectively.

\end{abstract}
  


\IEEEpeerreviewmaketitle       
\section{Introduction}
\label{sec:Intro}
With the rise of more advanced 3D displays, head-mounted displays and other advanced equipment, immersive media applications such as Free-viewpoint TV (FTV), 3DTV, and Virtual Reality (VR) have become hot topics for media ecosystems. FTV, which provides user with the `flying in the view' feeling by letting them navigate freely among different viewpoints, is one of the most popular scenarios in the area. In the FTV system, normally, only a limited set of input views are expected to be available and transmitted among all possible viewing angles that end user could select. As presented contents are usually synthesized using Depth-Image-Based Rendering technology (DIBR)~\cite{fehn2006interactive,fehn2004depth}, in addition to compression and smooth transition between views, reliable synthesis algorithms that are robust to sparser camera arrangements are critical factors with respect to the rendered quality. DIBR based algorithms have the tendency to introduce local non-uniform structure-related distortions. Following are some detailed introductions of the challenging spacial/temporal structure-related distortions that could be introduced by the FTV systems.


\textbf{Spatial structure-related distortions} within FTV system have the following characteristics: 1) non-uniform and locally distributed: unlike traditional global uniform artifact,~\textit{e.g.}, blocking artifacts, in most of the cases, the dominant spatial distortions of synthesized videos are the local non-uniform distortions and they distribute mostly around dis-occluded regions, which could be seen in the reference views but are occluded in the virtual views~\cite{bosc2011perceived}. These dis-occluded regions are commonly located at the  boundaries of objects, \textit{i.e.}, `regions of interest', and thus are more disturbing because local poor quality regions are with greater possibility to be perceived by observers than the global acceptable ones~\cite{ninassi2007does}; 2) structure-related local noncontinuous distortions are normally geometric distortions, which modify/deform the shape of the objects; 3) acceptable global shifting: DIBR based algorithms could also introduce global continuous shifting of objects. Observers are normally more tolerant to this type of distortion than the local serious one. Nevertheless, this type of distortion would be over-penalized by pixel to pixel metric like PSNR. 

\begin{figure}[]
\includegraphics[width=1\columnwidth]{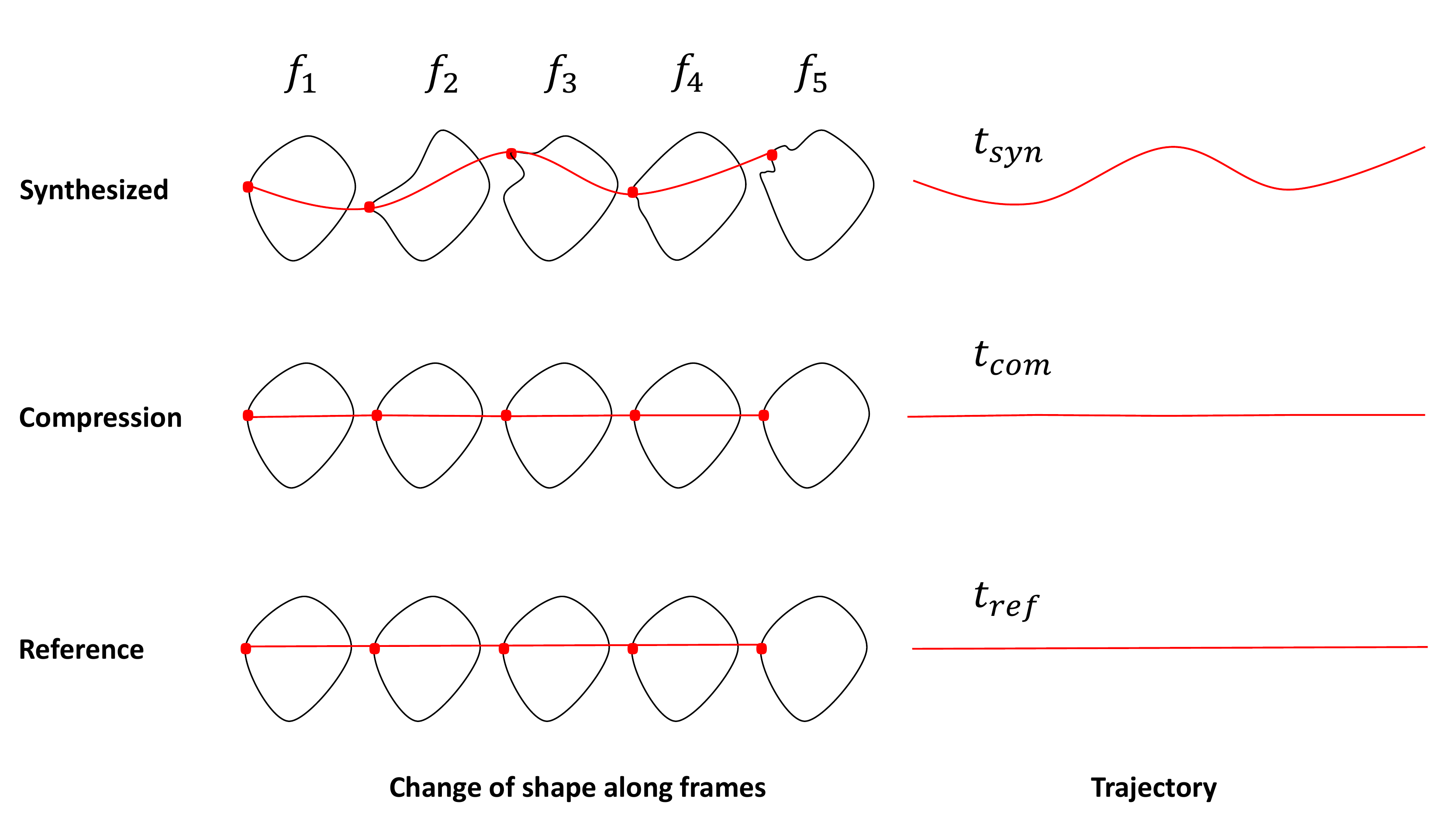}
  \caption{Example of temporal trajectory deformation caused by spatial geometric distortion. $t_{syn}$: trajectory from synthesized video; $t_{com}$: trajectory from video contain transitional compression artifacts; $t_{ref}$:  trajectory from reference video.}
   \label{fig:exa_Tem}
\end{figure}

\textbf{Temporal structure-related distortions} within FTV system could be categorized into two types.  1) temporal structure-related distortions within one viewpoint: considering each individual viewpoint, the spatial geometric distortion aroused by DIBR process will lead to temporal structure inconsistencies. Therefore, special temporal flickering in a form of structure (\textit{e.g.}, object boundaries) fluctuation could be observed within videos at a certain viewpoint location. For example, Fig.~\ref{fig:exa_Tem} shows the change of the shape of a static object along temporal axis, \textit{i.e.}, from the first frame $f_1$ to $f_5$. Different degrees of local structure-related distortions could be introduced differently among different viewpoints with different contents. 2) temporal structure-related distortions among viewpoints/unsmooth transition among viewpoints: considering the scenario of navigating among different viewpoints, local artifacts around dis-occluded regions, \textit{e.g.}, geometric distortions or inpainting related distortions, would incur structure inconsistencies from one view to another. The larger the baseline distance is used for synthesis, the more obvious the abrupt/sudden structural change could be observed when users switch their viewpoints from one to another. This unsmooth transition among different viewpoints could be considered as temporal flickering among viewpoints.  

In extreme cases, the entire viewing experience of one Free Viewpoint Video (FVV) can be ruined by only one severely distorted region in one synthesized view~\cite{gu2017model,ling2019prediction}. Quality assessment is vital for ensuring the quality of the entire system. Nevertheless, as most of existing image/video quality metrics have been tuned and designed to handle other types of distortions,~\textit{e.g.}, traditional uniform compression distortions including blocking artifact, blurriness~\textit{etc.}, they are mostly not suitable for FTV systems. New image/video quality assessment tools that can deal with these spatial and temporal structure-related distortions mentioned above are required. To evaluate the quality of free-viewpoint videos, some image and video quality assessment metrics have been proposed recently as introduced in the next section. However, their performances are limited, there is still a big room to improve. 

\section{Related Work}
\subsection{Limitations of Metrics Designed for Synthesized Views}
\label{sec:intro_QM}
\textbf{Image quality assessment metric designed for synthesized views}: in order to estimate the quality of synthesized views, there are many full reference (FR) metrics are proposed. The very first full reference approach that designed for evaluating the quality of synthesized images is proposed by Bosc~\textit{et al.}~\cite{bosc2011towards} by applying some prior knowledge acquired through subjective tests (\textit{e.g.}, the common localization of view-synthesis artifacts along contours) to SSIM. Following this idea, Conze~\textit{et al.}~\cite{conze2012objective} propose the view synthesis quality assessment (VSQA) metric, which improves SSIM with three visibility maps that characterizes the complexity of the images. Later, the `3D synthesized view image quality metric' (3DswIM) is proposed by Battisti~\textit{et al.}~\cite{battisti2015objective}. This metric is based on statistical features of wavelet sub-bands. In addition, Tsai and Hang~\cite{tsai2013quality} propose a metric based on compensating the shifts of the objects that appear in synthesized views by calculating the noise around them. Considering the fact that using multi-resolution approaches could increase the performance of image quality metrics, Sandi{\'c}-Stankovi{\'c}~\textit{et al.} develop the `Morphological Wavelet PSNR' (MW-PSNR) using a morphological wavelet decomposition~\cite{sandic2015dibr}. Later they extend the work by using a multi-scale decomposition based on morphological pyramids, which is called `Morphological Pyramid PSNR' (MP-PSNR)~\cite{sandic2015dibrMP}. Recently, Stankovi{\'c}~\textit{et.al.}~\cite{sandic2016dibr} point out that PSNR is more consistent with human judgment when it is calculated at higher morphological decomposition scales. They thus proposed reduced versions of the morphological multi-scale measures called reduced MP-PSNR and reduced MW-PSNR correspondingly (denoted as MP-PSNR$_r$ and MW-PSNR$_r$). According to their experimental results, the reduced versions (\textit{i.e.}, MP-PSNR$_r$ and MW-PSNR$_r$) outperform the full versions. Li~\textit{et al.}~\cite{li2018quality} propose LOGs by considering both the geometric distortions as well as the sharpness of the images. NIQSV+~\cite{tian2018niqsv+} is proposed based on a strong hypothesis that high-quality images are consist of flat areas separated by edges. Another state-of-the-art image quality metric is the EM-IQM (\textit{i.e.}, EM$_{spa}$ in this paper) in \cite{EM-IQA} that quantifies the spatial structure deformations using elastic metric. 

All the image metrics mentioned above suffers from at least one of the drawbacks mentioned below: 1) The human visual system is sensitive to severe local artifacts~\cite{moorthy2009visual,ninassi2007does}. The most upsetting artifacts in synthesized images are the inconsistent local geometric distortions instead of the consistent global uniform distortions. However, most of the existing metrics process the entire image equally and thus fail to locate and quantify local geometric distortions properly. Sensitive region selection should be considered as a pre-process module to select regions with structure-related distortions. 2) Global shifting within certain limits is acceptable for human observers but is punished severely by point-to-point based metrics. Due to equal-weighted pooling and point-wise comparison, some image quality assessment metrics mistakenly emphasize the consistent global shifting artifacts. 3) All of these metrics are not capable of quantifying the amount of temporal structure-related distortions.

\textbf{Video quality assessment metric designed for synthesized views}:   The `Peak Signal to Perceptible Temporal Noise Ratio' (PSPTNR) metric, introduced by Zhao and Yu~\cite{zhao2010perceptual}, quantifies temporal artifacts that can be perceived by observers in the background regions of the synthesized videos. Similarly, Ekmekcioglu~\textit{et al.}~\cite{ekmekcioglu2010depth} propose a video quality metric by using depth and motion information to locate the degradations. The state-of-the-art video metric designed for free viewpoint videos is recently introduced by Liu~\textit{et al.}~\cite{liu2015subjective}. Their proposed metric (Liu-VQM) considers the spatio-temporal activity and the temporal flickering that appears in synthesized video sequences. However, none of the aforementioned video quality metrics is designed to quantify the unsmooth transition among views (temporal structure inconsistency observed during view switch). Compared to temporal distortions that could be observed at one viewpoint, during navigation, structure-related distortions could be amplified and new temporal structure deformation could be noticed. Therefore, temporal structure-related distortions are more challenging and should not be underestimated.

In summary, a new video quality metric, which is able to quantify the aforementioned spatial-temporal structure-related distortions, is in urgent need. 

\subsection{So, How to Better Quantify the Spatial-Temporal Structure-related Distortions? } 
\begin{figure}[ ]
\centering
\subfloat[Reference frame ]{\label{ex0}
\includegraphics[width=0.33\columnwidth]{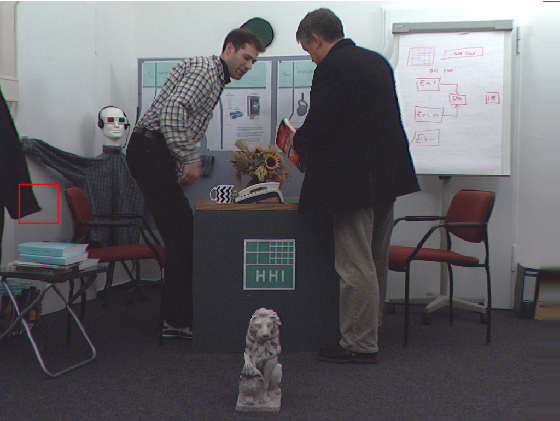}}
\subfloat[Synthesized frame~\cite{telea2004image}]{\label{ex1}
\includegraphics[width=0.33\columnwidth]{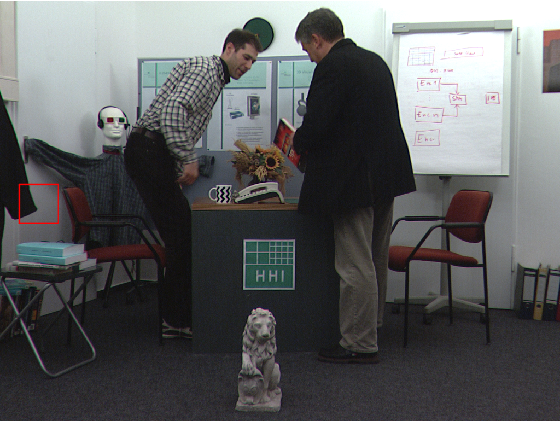}}
 \subfloat[Synthesized frame~\cite{ndjiki2010depth}]{\label{ex1}
\includegraphics[width=0.33\columnwidth]{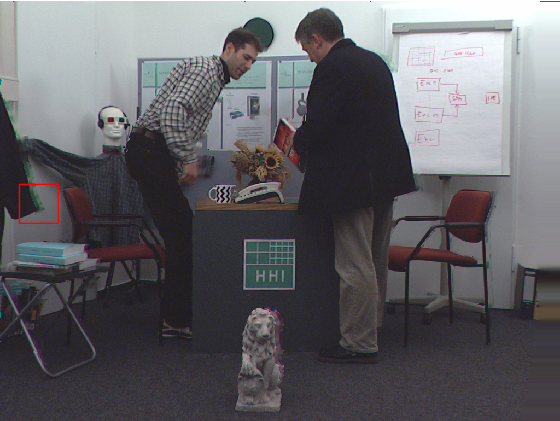}} \\

\subfloat[Reference patch ]{\label{ex0}
\includegraphics[width=0.25\columnwidth]{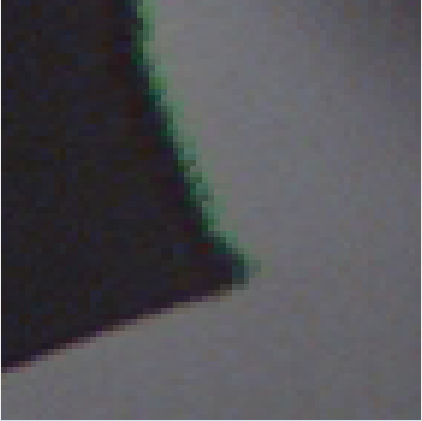}}
\subfloat[PSNR=22.2077]{\label{ex1}
\includegraphics[width=0.25\columnwidth]{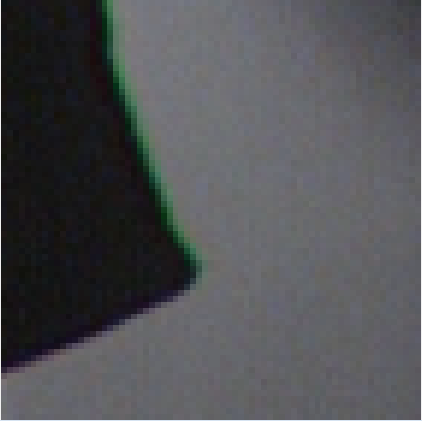}}
 \subfloat[PSNR=28.1877]{\label{ex1}
\includegraphics[width=0.25\columnwidth]{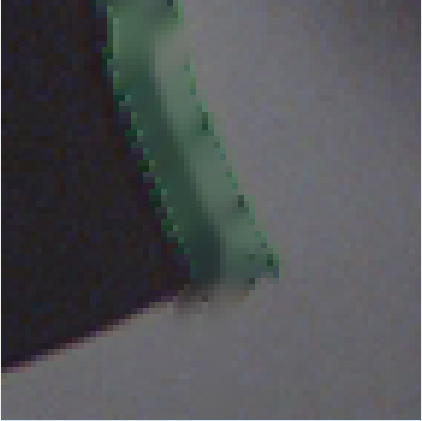}} \\

\subfloat[Reference patch ]{\label{ex0}
\includegraphics[width=0.25\columnwidth]{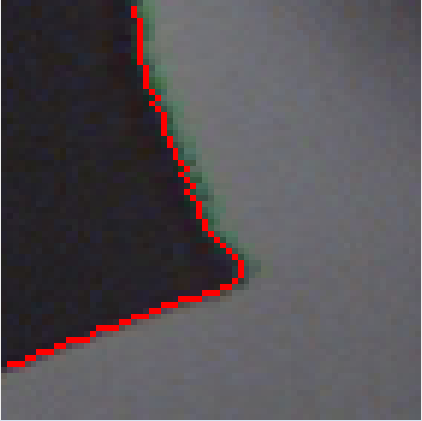}}
\subfloat[D$_{EM}$=0.0894]{\label{ex1}
\includegraphics[width=0.25\columnwidth]{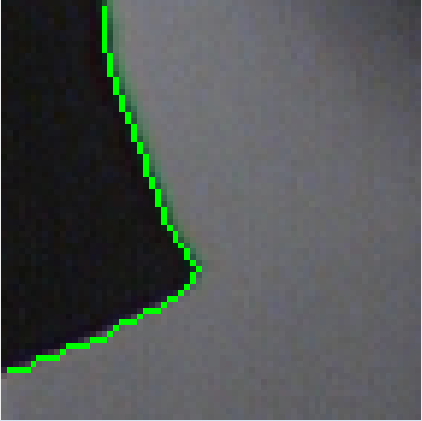}}
 \subfloat[D$_{EM}$=0.3068 ]{\label{ex1}
\includegraphics[width=0.25\columnwidth]{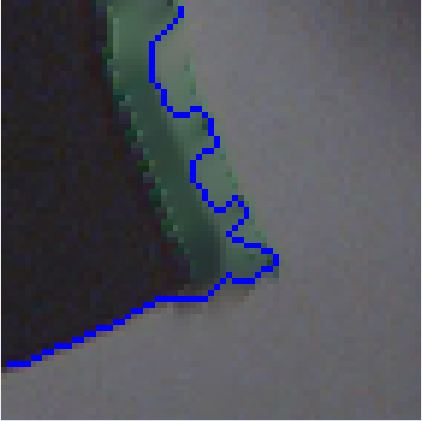}} \\

\caption{Examples of advantages of using elastic metric compared to pixel-to-pixel metric PSNR. Rows  (from up to down): Reference/Synthesized images; Patches from images (a)-(c) bounded by red bounding boxes; Extracted contours of patches (d)-(f). Columns (from left to right): reference image, synthesized image with synthesis algorithm proposed in~\cite{telea2004image}, synthesized image with synthesis algorithm proposed in~\cite{ndjiki2010depth}. }
\label{fig.example_SEM}
\end{figure}

\textbf{Elastic Metric (EM)}, which is first proposed in~\cite{mio2007shape}, is capable of quantifying the deformation between curves and thus could be the solution for quantifying the aforementioned specific distortions. It is first utilized in our previous work~\cite{EM-IQA} to evaluate the quality of synthesized views spatially by calculating the amount of stretching or bending between curves/shapes in the reference and synthesized frames. Example of the advantage of using elastic metric compared to PSNR is shown in Fig.~\ref{fig.example_SEM}. By checking patches in Fig.~\ref{fig.example_SEM} (e) and (f), it is obvious that the patch in Fig.~\ref{fig.example_SEM} (e) with the slightly shifted object is of better visual quality than the one in Fig.~\ref{fig.example_SEM} (f) with obvious structure-related distortions (\textit{i.e.}, ghosting artifact). However, PSNR (the higher the score the better the quality) incorrectly indicates that the quality of (f) is better than (e). In the contrary, elastic metric (\textit{i.e.}, $D_{EM}$, the higher the scores the the larger the amount of deformations between two compared curves) accurately points out that the blue curve in Fig.~\ref{fig.example_SEM} (i) is more severely deformed compared to the red curve in Fig.~\ref{fig.example_SEM} (g), indicating worse quality. After being able to quantify the spatial temporal structure-related distortions, then how to quantify the temporal ones, especially the ones observed during view switch? In this paper, multi-scale trajectory is employed along with elastic metric to handle this challenging problem.

\textbf{Motion} plays a vital role in visual perception of the contents and the perceived quality of sequence since 1) clues related to the objects' shapes are provided; 2) in most cases, visual attention is tend to be drawn on moving objects~\cite{wandell1995foundations,itti2001computational,seshadrinathan2010motion}. Human Visual System (HVS) tends to trace the salient moving objects when viewing a sequence \cite{todd1981visual}, thus distortions around the moving objects may attract greater attention from the observers. There are already quality metrics designed based on this phenomenon \cite{wang2007video,barkowsky2009temporal}. In the case of free view-point videos generated with synthesized views, apart from the special spatial distortions as described in~\cite{battisti2015objective}, the synthesized videos suffer also from special temporal degradation related to motions of objects within one viewpoints or navigations among different viewpoints~\cite{liu2015subjective} as summarized in the previous section. Therefore, the success of one video metric dedicating to ensuring the quality of the entire FTV system relies on its capability of modeling and accounting both structure and motion perception in the HVS.

\textbf{Motion trajectory}, which traces moving objects, provides human observers with important spatial-temporal information for perceiving/detecting moving objects,~\textit{e.g.,}velocity, direction and even spatial information of the objects~\cite{krekelberg1999temporal}. Since the detectability of a moving object could be impacted by the structural motion information (which could be represented by the motion trajectories), deformation of trajectory or changes of structure information along the trajectory that caused by distortions may affect the way how human detect objects and thus affect the perceived quality. Hence, on one hand, considering the characteristics of distortions produced by DIBR processes, synthesized sequences could be represented in sets of trajectories and their perceived quality could be evaluated with the trajectories and neighborhoods along them. For example, as shown in Fig. \ref{fig:exa_Tem}, due to the change of shape of  the object, the trajectory of the synthesized sequence $t_{syn}$ that traces one of the key point on the shape is deformed compared with the one of the reference sequence $t_{ref}$ , while the one of the sequence that contain only common compression artifact remains almost unchanged. If one could quantify the amount of deformation of trajectories caused by related processes, the quality of the synthesized videos could be indicated. Since motion trajectories within sequence could be considered as open-curves, elastic metric is of potential to be used as a measure to quantify the deformation between the trajectory in a synthesized sequence and the one in the original sequence. On the other hand, as spatial-temporal distortions mainly happen around dis-occluded regions and distortions within the regions of moving objects are less tolerant for observers, the process of detecting meaningful moving trajectories could serve as a way to select severe distorted regions.


\textbf{Multi-scale approaches}, which transfer signals into a form of multi-scale representation, could be used to quantify structure loss caused by synthesized related artifacts.
On one hand, the perceivability of videos' details is decided not only by the observer's visual system, but also by the viewing conditions (e.g., display resolution and viewing distance) \cite{ms-ssim} Therefore, it is more reasonable to employ multi-scale strategy for quality assessment to take different subjective factors and configurations into account as claimed in~\cite{felzenszwalb2008discriminatively}. On the another hand, as pointed out in~\cite{ms-ssim}, vision at a glance reflects high-level mechanism. Observers normally obtain the structure of the content first before looking into the details with scrutiny. In another word, structure of one content could be obtained from a lower resolution, while the details could be obtained through a higher resolution. The concept of multi-scale strategy is in line with human visual mechanism. Therefore, dissimilarity between the synthesis and reference videos calculated in different scales could be used differently depends on the subjective configurations/parameters to improve quality assessment metrics.

Based on the discussion above, in order to better evaluate the quality of free view-point videos by considering the characteristics of the spatial-temporal structure related distortions, an elastic metric and multi-scale trajectory based video quality assessment metric (EM-VQM) is proposed in this paper. The contributions of this paper are three-folds: 1) multi-scale motion trajectory is used as a proxy for temporal sensitive regions selection;  2) elastic metric is used to quantified the amount of motion trajectory deformations; 3) motion-structure-related descriptors are extracted along the multi-scale motion trajectories and used to quantify the spatial-temporal structural dissimilarities between the reference and synthesized videos. 

The remainder of this paper is organized as follows. In Section~\ref{sec:PM}, the proposed model is introduced in detail. Then, the experimental results and analysis are presented in Section~\ref{sec:exp}. Finally, conclusions are given in Section~\ref{sec:con}.

\section{The Proposed Model}
\label{sec:PM}
\begin{figure}[ ]
 \includegraphics[width=1\columnwidth]{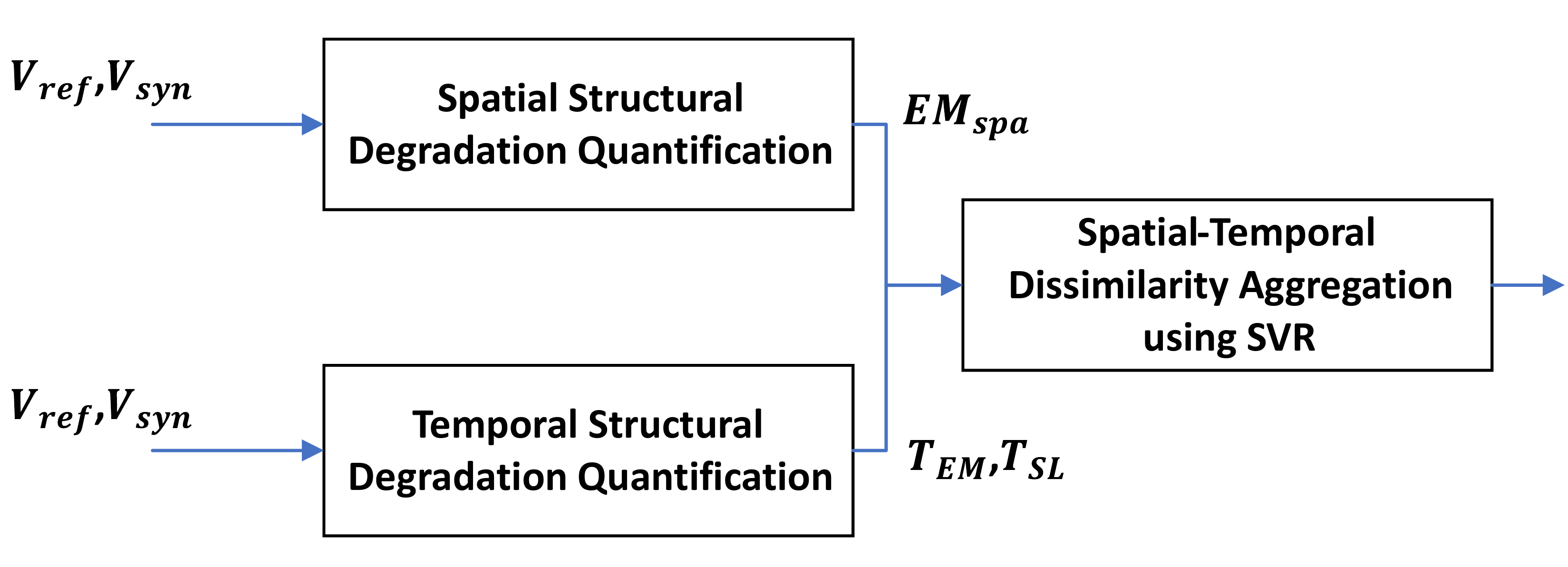}
  \caption{Overall framework of the proposed EM-VQM metric.}
   \label{fig:overall_framework}
\end{figure}

 The proposed elastic metric based video quality assessment metric is composed of two parts, including one part for quantifying the spatial structural degradation (Section~\ref{sec:SSD}) and another part for quantifying temporal structural degradation (Section~\ref{sec:TSD}). After computing the amount of spatial and temporal structure-related distortions at multi-scales separately, they are aggregated (Section~\ref{sec:STA}) to predict the overall quality score of one synthesized videos as illustrated in Fig.~\ref{fig:overall_framework}.

\subsection{ Quantify Spatial Structural Degradation}
\label{sec:SSD}
\subsubsection{Spatial sensitive regions selection using key point matching}
\label{ssec:LRS}

\begin{figure}[!htbp]
 \includegraphics[width=1\columnwidth]{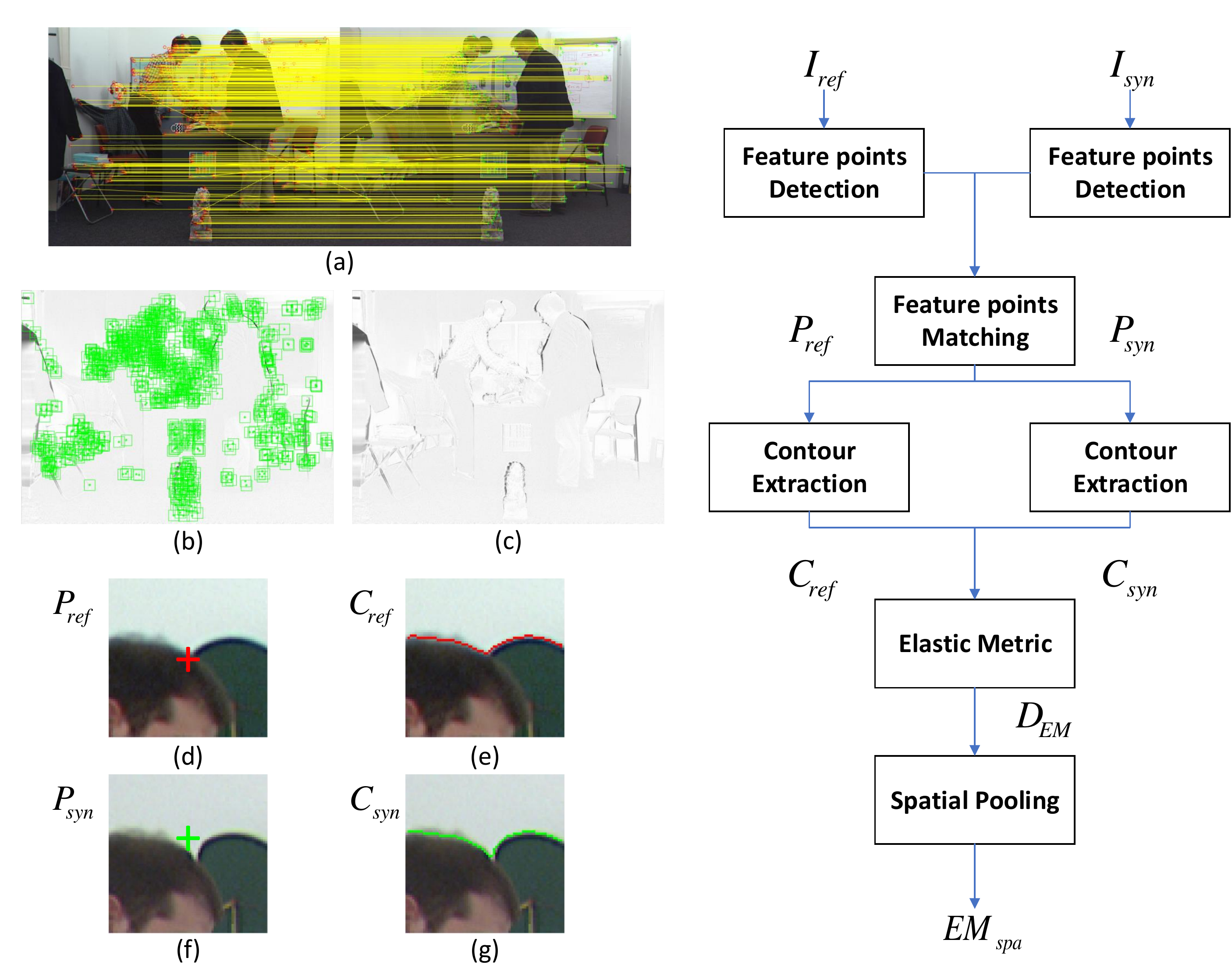}
  \caption{Diagram of spatial structural degradation quantification. (a) Example of SURF feature points matching, where matched SURF feature points within reference and synthesized frames are connected with yellow color lines.  (b) Selected sensitive patches, \textit{i.e.}, patches centering at the matched SURF feature points bounded by green color boxes, plotted on the error map. (c) Error map obtained by comparing the reference and the synthesized frames (the darker the color the more severe the distortions within the regions). (d) Example of one sensitive region (patch) in the reference frame, whose center is one SURF feature point labeled with a red color cross. (e) Contour extracted from the patch. (f) Example of one sensitive region (patch) in the synthesized frame, whose center is one SURF feature point labeled with a green cross. (g) Contour extracted from the patch. } 
   \label{fig:spatial_framework}
\end{figure} 
 
Unlike traditional distortions, which scatter over the entire frames, synthesized distortions distribute sparsely/locally and thus need to be selected and dealt with particularly. Structural descriptor Speeded Up Robust Features (SURF)~\cite{bay2006surf} is of the ability to detect important structure key points within images/videos. Since local synthesized artifacts are likely to appear around these key-points regions and draw greater attention from the human observers, SURF point detection and matching is used for sensitive regions selection, where geometric distortions are less acceptable for observers. It is worth mentioning that, SURF points matching could also compensate the global objects shifting artifacts and avoid over penalizing the acceptable uniform artifact as described and confirmed in our previous work~\cite{EM-IQA}. 

\subsubsection{Curve extraction based on patch segmentation} 
\label{ssec: CE_PS}
In order to check the magnitudes/amount of spatial-structural deformations after synthesis using elastic metric, curves need to be first extracted in an efficient way. After the process of local regions selection and matching,  the SLIC superpixel approach~\cite{achanta2012slic} is then used to segment the matched patches $P_{ori}$ and  $P_{syn}$ centering at matched key points. Then, the boundaries of the segmented super-pixels set  $SP_{ori}$ and $SP_{syn}$  are extracted as the closed curves, which will be proceeded for latter comparison. Afterwards, the fast superpixels matching algorithm proposed in our previous work~\cite{EM-IQA} is used to further obtain the set matched closed curves $( C_{ori},C_{syn} )$. 

\subsubsection{Spatial curves comparison using elastic metric}
\label{ssec:CC based on EM in ES}
The amount of spatial structural deformation is computed by comparing each matched closed curves $(c_{ori}^{i} , c_{syn}^{j}) \subset {( C_{ori},C_{syn} )}$ obtained in the previous subsection using the elastic metric proposed in~\cite{mio2007shape,srivastava2011shape}. 
A curve is first parameterized as $c$ with $k\in{K}$ as the parameter. It is defined as 
\begin{equation} 
	c: K \rightarrow (x,y) \in{\mathbb{R}^n},
	\label{curve representation}
\end{equation}
where $(x,y)$ represents the the coordinates of each point of curve. In general, $K=[0,1]$. For closed curves, $K=\mathbb{S}^1$. The parameterized curve could be then represented using the Square-Root Velocity (SRV) function defined as $ q : K \rightarrow (x,y) \in{\mathbb{R}^n} $, where 
\begin{equation} 
	q(k)\equiv F(\dot{c}(k))= \dot{c}(k)/ \sqrt{\| \dot{c}(k)\|} ,
	\label{srv function}
\end{equation}
where $\|\cdot\|$ is the Euclidean 2-norm in $\mathbb{R}^n$ and 
$\dot{c}=  \frac{dc}{dk} $. The original curve could be derived reversely with the following equation
\begin{equation} 
c(k)= \int^k_0 q(s) \|q(s)\|ds.
\end{equation}
Later, $\phi:K \rightarrow \mathbb{R}$, with $\phi(k)=ln(\|\dot{c}(k)\|)$ and $\theta :K \rightarrow \mathbb{S}^{n-1}$, with $\theta(k)=\dot{c}(k)/\|\dot{c}(k)\|$ are further defined by Srivastava \textit{et al.} in~\cite{srivastava2011shape} to quantify curve deformations. Based on these definition, a riemannian metric named `Elastic Metric' $D_{EM}$ on the tangent space $\tau \ $　of $ \Phi \times \Theta$ could be then defined based on the computation of the following inner product:
\begin{equation} 
	\begin{split}
		& D_{EM}=\langle (u_1,v_1),(u_2,v_2) \rangle_{(\phi,\theta)}\\
		&= a^2\int_Du_1(k)u_2(k)e^{\phi(k)}dk
		+b^2\int_Dv_1(k)v_2(k)e^{\phi(k)}dk,
	\end{split}
	\label{elastic metric}
\end{equation}
where $\langle \cdot \rangle$ denotes the standard dot product in $\mathbb{R}^n$ and $(u_1,v_1),$ $(u_2,v_2)$ $\in {\tau_{\phi , \theta}(\Phi \times \Theta)}$. As explained in \cite{mio2007shape,srivastava2011shape}, $u_1$ and $u_2$ in the first integral are variations of the log speed $\phi$ of the curves, while $v_1$ and $v_2$ in the second integral are the variations of the direction $\theta$ of the curves. The first and second integrals could be interpreted to measure the amount of `stretching' and `bending' correspondingly and $a^2$,$b^2$ are two parameters chosen to penalize these two types of deformations. To calculate Eq.~\eqref{elastic metric} more efficiently, the SRV formulationin Eq.~\eqref{srv function} are used and adjusted in terms of $(\phi , \theta)$  by defining $q(k)=e^{\frac{1}{2}\phi(k)}\theta(k)$. Afterwards, the tangent  vectors to $\mathbb{L}^2(K,\mathbb{R}^n)$ at q is obtained with
$r=\frac{1}{2}e^{\frac{1}{2}\phi}u\theta+e^{\frac{1}{2}\phi}v$. For two elements $r_1$ and $r_2$ of ${\tau_{\phi , \theta}(\Phi \times \Theta)}$, computing the $\mathbb{L}^2$-metric (elastic metric) of them yields
\begin{equation} 
	\begin{split}
		&D_{EM}(c_{ori}^{i} , c_{syn}^{j})  =  \langle r_1,r_2 \rangle \\
        & = \int_K{\langle \frac{1}{2}e^{\frac{1}{2}\phi}u_1\theta+e^{\frac{1}{2}\phi}v_1,\frac{1}{2}e^{\frac{1}{2}\phi}u_1\theta+e^{\frac{1}{2}\phi}v_2 \rangle}dk\\
		&=\int_K{(\frac{1}{4}e^\theta u_1u_2+e^\theta\langle v_1,v_2 \rangle)}dk.
	\end{split}
	\label{final elastic metric}
\end{equation}

\subsubsection{Spatial Pooling}
\label{ssec:SP}
The D$_{EM}$ calculates local elastic dissimilarity between each pair of matched closed curves from the reference and synthesized images based on region selection and elastic metric described in the previous section. As discussed in previous sections, human observers tend to perceive `poor' regions than the `good' ones within an image. For DIBR based synthesized images, the sensitive disoccluded regions are the `poor' regions and should be penalized during the quality assessment. 

As the curves are only extracted from the selected regions, where the annoying local distortions mainly appear, there is no need to apply other specific pooling strategies for pooling the elastic dissimilarity scores. Moreover, due to local regions selection, artifacts in local important disoccluded regions are penalized sufficiently, and at the same time, the global consistent artifacts are not over penalized. Hence, the final objective score is calculated by simply summing out all the elastic dissimilarities values, which is defined as
\begin{equation}
	EM_{spa}= \sum D_{EM}(  c_{ori}^{i} , c_{syn}^{j} ),
\end{equation}
 where $ (c_{ori}^{i} , c_{syn}^{j}) \subset {( C_{ori},C_{syn} )}$.

\subsection{Quantify Temporal Structural Degradation }
\label{sec:TSD}
\begin{figure}[ ]
\centering
 \includegraphics[width=0.7\columnwidth]{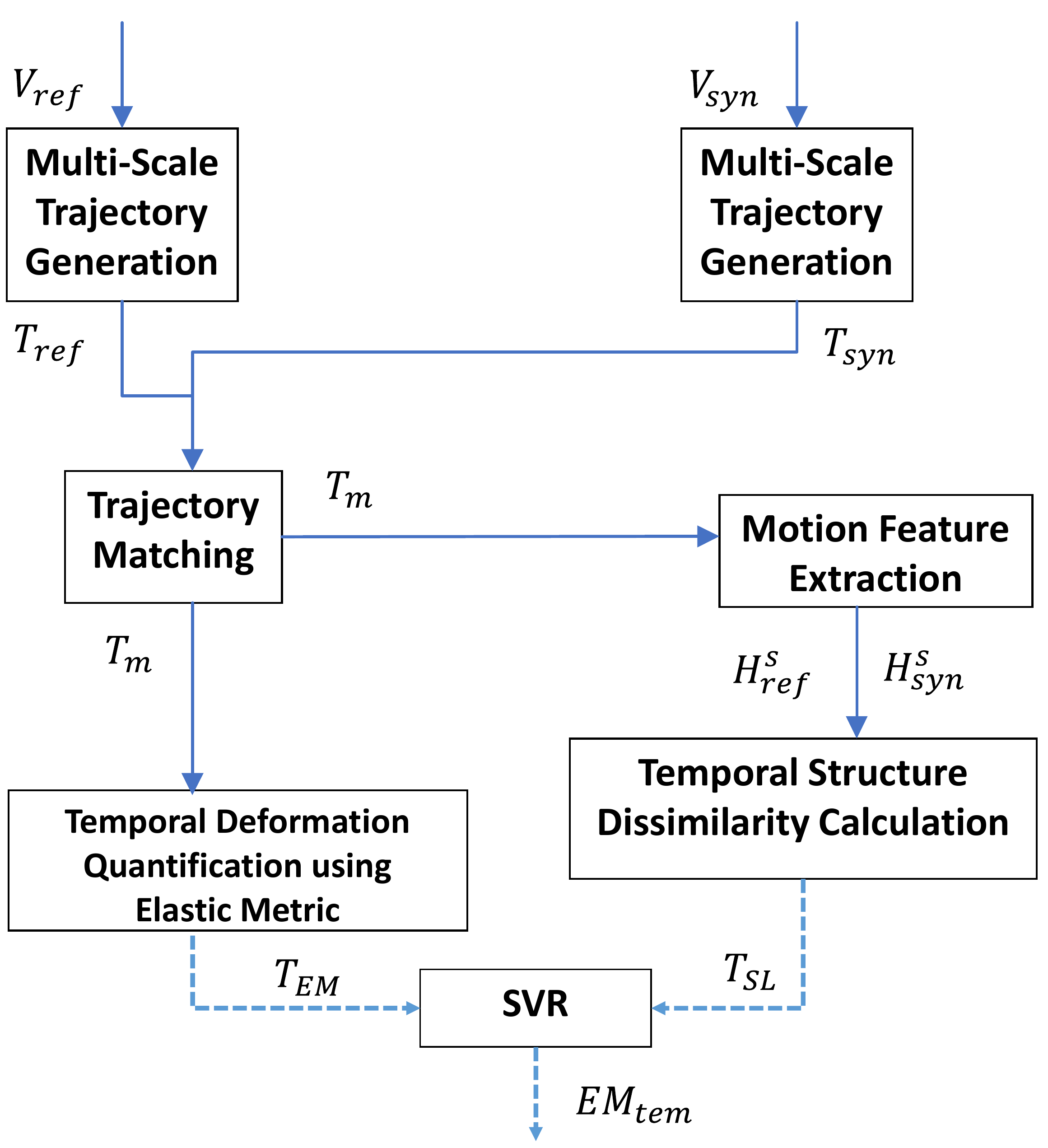}
  \caption{Framework of temporal structural degradation quantification.}
   \label{fig.frame_work_T}
\end{figure}

In this section, details of how to quantify the non-uniform temporal structure-related distortions are given. The framework is summarized in Fig.~\ref{fig.frame_work_T}. As motion trajectory reveals important structural-motion information, distortions along motion trajectory are thus easier to be noticed by observers. Based on this fact, multi-scale trajectory representation is exploited in this work to quantify local structure related distortions that affect the quality of the synthesized sequence. In the proposed scheme, given one synthesized video $V_{syn}$ and its reference video $V_{ref}$, they are firstly represented as a set of multi-scale trajectories $T^s_{syn}$ and $V^s_{ref}$ respectively (\textit{i.e.} trajectory at different scales), where $s$ indicates a certain scale. Considering the characteristics of DIBR based synthesis techniques, the neighborhoods around the trajectories could be considered as the candidates regions, where local non-uniform distortion may appear and severely degrade the quality of the entire sequence. With the multi-scale trajectory representation, spatial-temporal structure-related features, in the form of histograms,~\textit{i.e.}, $H^s_{syn}$ and $H^s_{ref}$, along the trajectories are extracted. Finally, deformations of the object structures in the form of deformations of trajectories, \textit{i.e.} T$_{EM}$, could be quantified using elastic metric with $T^s_{syn}$ and $T^s_{ref}$, while the structural losses along trajectories, \textit{i.e.} $T_{SL}$, could be quantified with the temporal-structure features/histograms $H^s_{syn}$ and $H^s_{ref}$.

To check the performance of the proposed temporal structural distortion estimator, the obtained T$_{EM}$, T$_{SL}$ are combined as a new video quality assessment metric, which is denoted as EM$_{tem}$. It has to be emphasized that, the final proposed EM-vQM is obtained by integrating T$_{EM}$, T$_{SL}$, and EM$_{spa}$ using SVR as illustrated in Fig.~\ref{fig:overall_framework} instead of combing EM$_{spa}$ and EM$_{tmp}$. Details of the computation of spatial-temporal structural dissimilarity between a synthesized sequence and its reference, \textit{i.e.}, T$_{EM}$, T$_{SL}$, are given below.

\begin{figure*}[!htbp]
\centering
\subfloat[ ]{\label{ex0}
\includegraphics[width=0.4\textwidth]{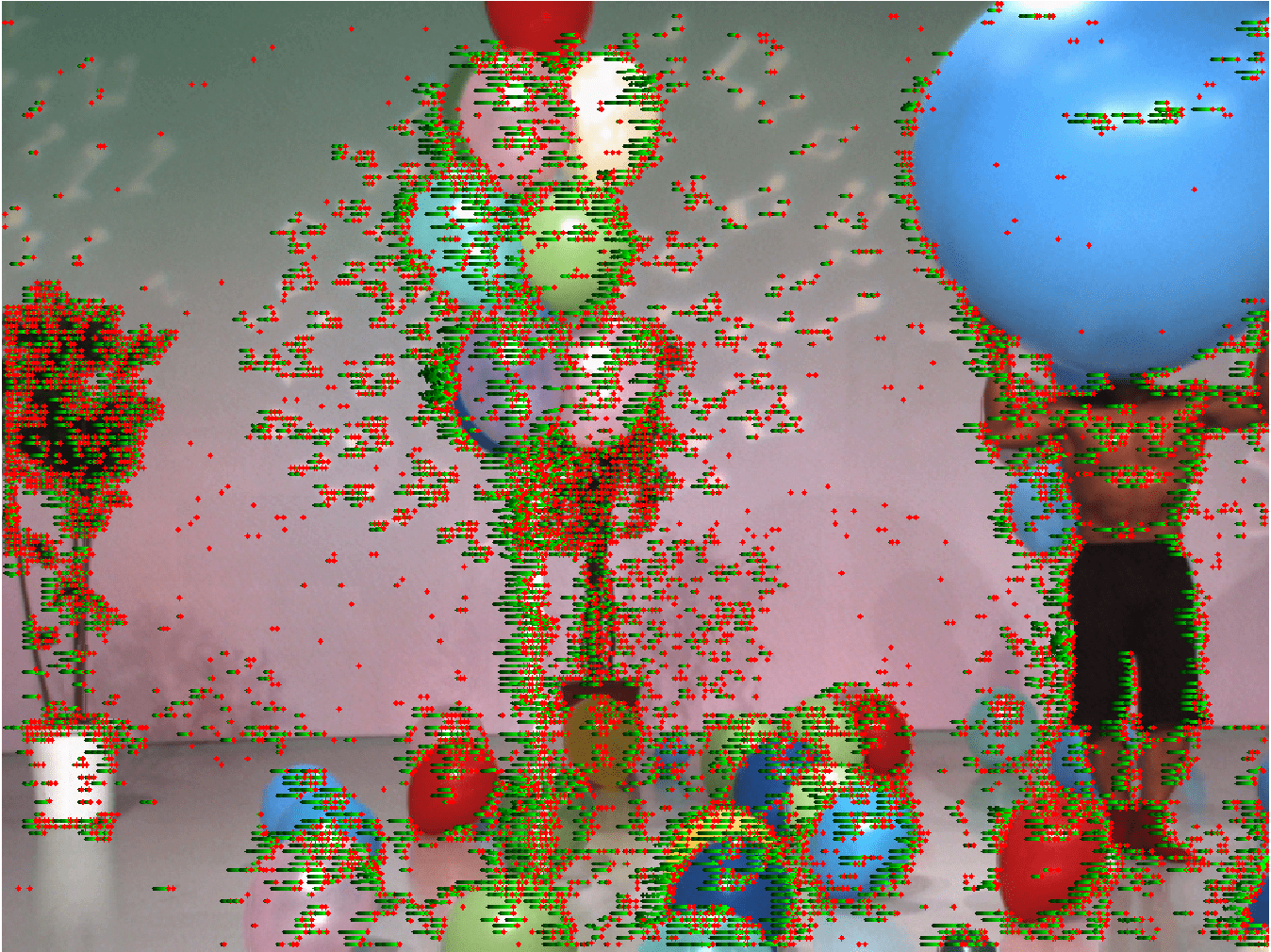}}
\subfloat[ ]{\label{ex1}
\includegraphics[width=0.4\textwidth]{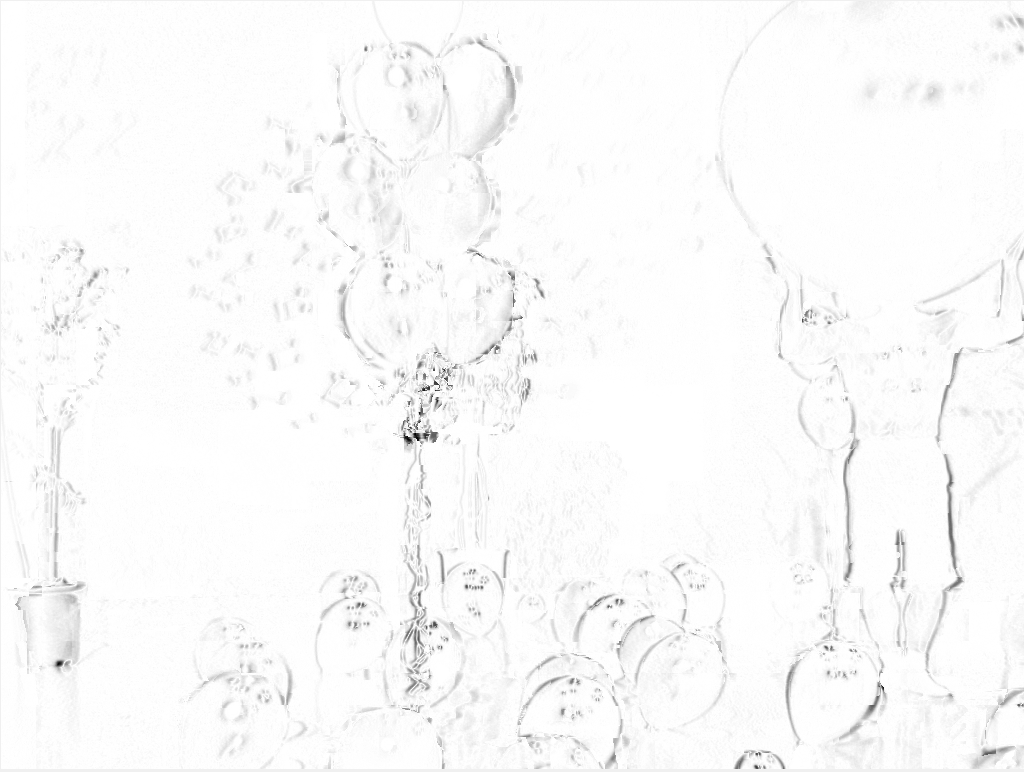}} 
\caption{Example of temporal sensitive regions selection: (a) Example of dense motion trajectory, where red points represents the key points in the current frame and the green lines connect the key points at the current frames with the ones in the previous frame. (b) Error map between frames extracted from the reference and synthesized views, where the darker color the more severe the distortions are within the regions.}
\label{fig:example_TRA}
\end{figure*}

\begin{figure}[ ]
\centering
 \includegraphics[width=1\columnwidth]{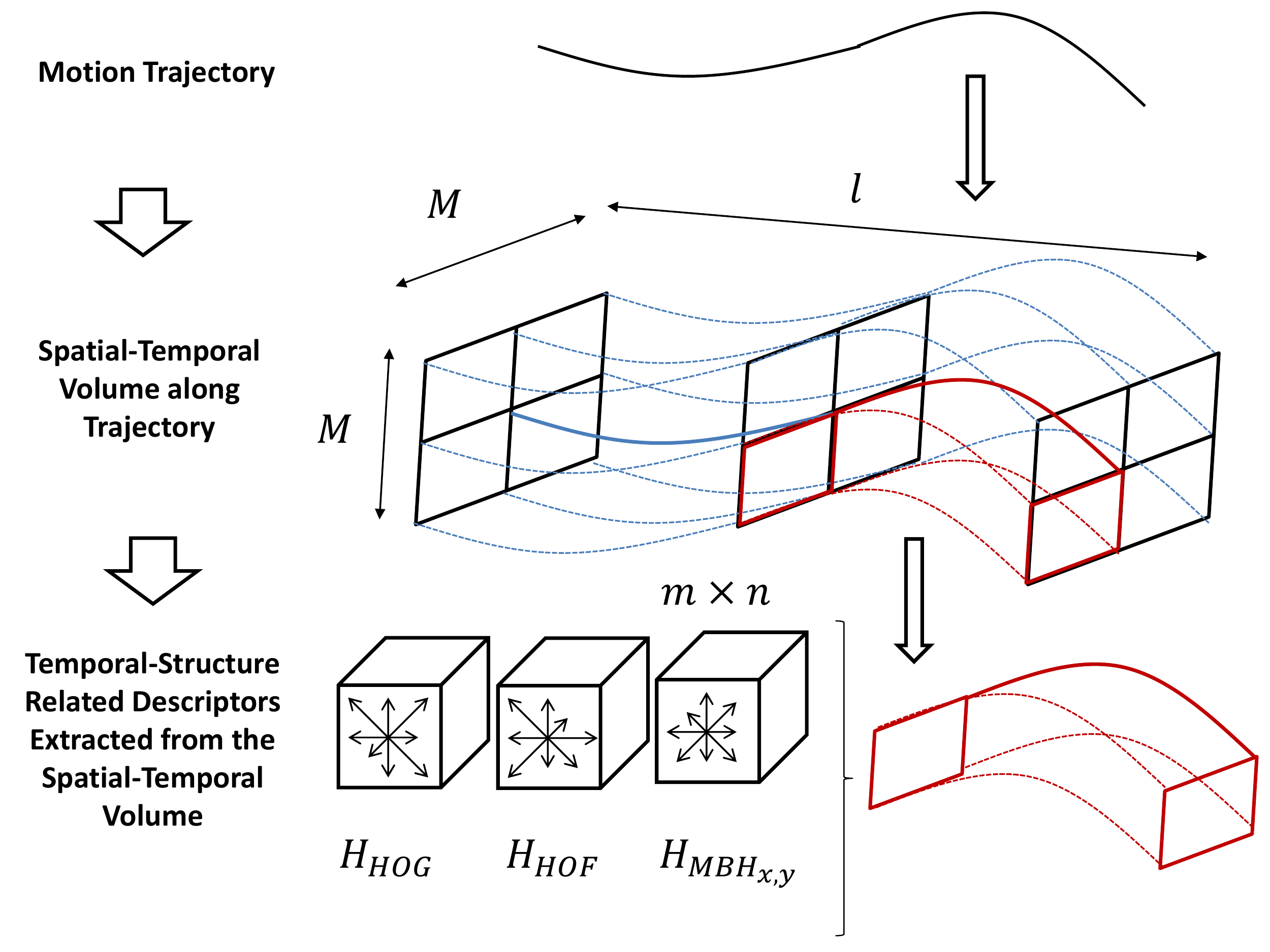}
  \caption{Illustration of temporal-structure related trajectory descriptors extraction along motion trajectories.}
   \label{fig.frame_mf}
\end{figure}

\subsubsection{Multi-scale motion trajectory representation as spatial-temporal distortion regions selection}
Dense motions trajectory, which is first proposed in~\cite{wang2013dense} by Wang~\textit{et al.}, is first utilized in this work to represent free-viewpoint videos. It is a spatial-temporal representation for video with multi-scale dense trajectories and descriptors of structural-motion boundary along the trajectories. 

After generating the multi-scales version of the video $V$ with $S$ spatial scales, feature points are sampled on each spatial scale $s \in \{1,...,S\} $ with a sampling step $W$. In this work, seven scales are considered. Each spatial scale increases with a factor of $1 / \sqrt[]{2}$ as done in~\cite{wang2013dense}. Considering that most of the local disturbing geometric distortions are commonly located around the boundaries of the objects (\textit{i.e.}, the dis-occluded regions) instead of homogeneous texture regions, points within regions that do not contain any structure are removed. Later, sampled points on each spatial scales are then tracked by using Large Displacement Optical Flow algorithm (LDOF) proposed in~\cite{brox2011large}. Then, each trajectory $t_s$ obtained in a certain scale $s$ could be represented as a sequence of points $  (p_1, ...,p_{f_i}, ... , p_l )$ with a length of $l$ ($l$ is set as 15 in this paper). In $t_s$, $p_{f_i}$ is a feature point at frame $f_i$, which is spatially-temporally related to features points in previous and latter frames, \textit{i.e.}, $p_{f_{i-1}}$ and $p_{f_{i+1}}$, according to the calculated optical components.  As human observers are more sensitive to moving structural regions, \textit{e.g.}, moving objects, static trajectories that do not contain any motion are pruned. It is worth mentioning that the process of generating trajectories could be served as a proxy to select candidate sensitive regions, as local synthesized distortions that distribute around the moving objects attract most of the attention from the observers. Example is shown in Fig.~\ref{fig:example_TRA}, where it can be observed that most of the error regions have been covered by the detected motion trajectories.

\subsubsection{Temporal-structure-related trajectory descriptor}
In order to better quantify the changes of spatial-temporal structural information along trajectories due to synthesis process, with respect to the reference, three motion-structure related descriptors~\cite{brox2011large} are extracted for each trajectory. More specifically, they are Histogram of Oriented gradient (HOG)~\cite{wang2009evaluation}, Histogram of Optical Flow (HOF) \cite{laptev2008learning} and Motion Boundary Histogram (MBH)~\cite{dalal2006human} extracted within a spatial-temporal volume that is aligned with one trajectory $T_s$ as illustrated in Fig.~\ref{fig.frame_mf}. MBH is computed with the the derivatives of both the horizontal and vertical elements of one optical flow, which further ends up into two histograms for each component as $MBH_h$ and $MBH_v$ normalized with $L_2$ norm. Therefore, for each trajectory at a scale $s$, four spatial-temporal structural histograms, including $H^s_{HOG}$, $H^s_{HOF}$, $H^s_{MBHx}$ and $H^s_{MBHy}$, are obtained after feature extraction procedure.

\begin{figure*}[ ]
\centering
  \begin{minipage}[b]{0.4\linewidth}
    \centering
    \includegraphics[width=\linewidth]{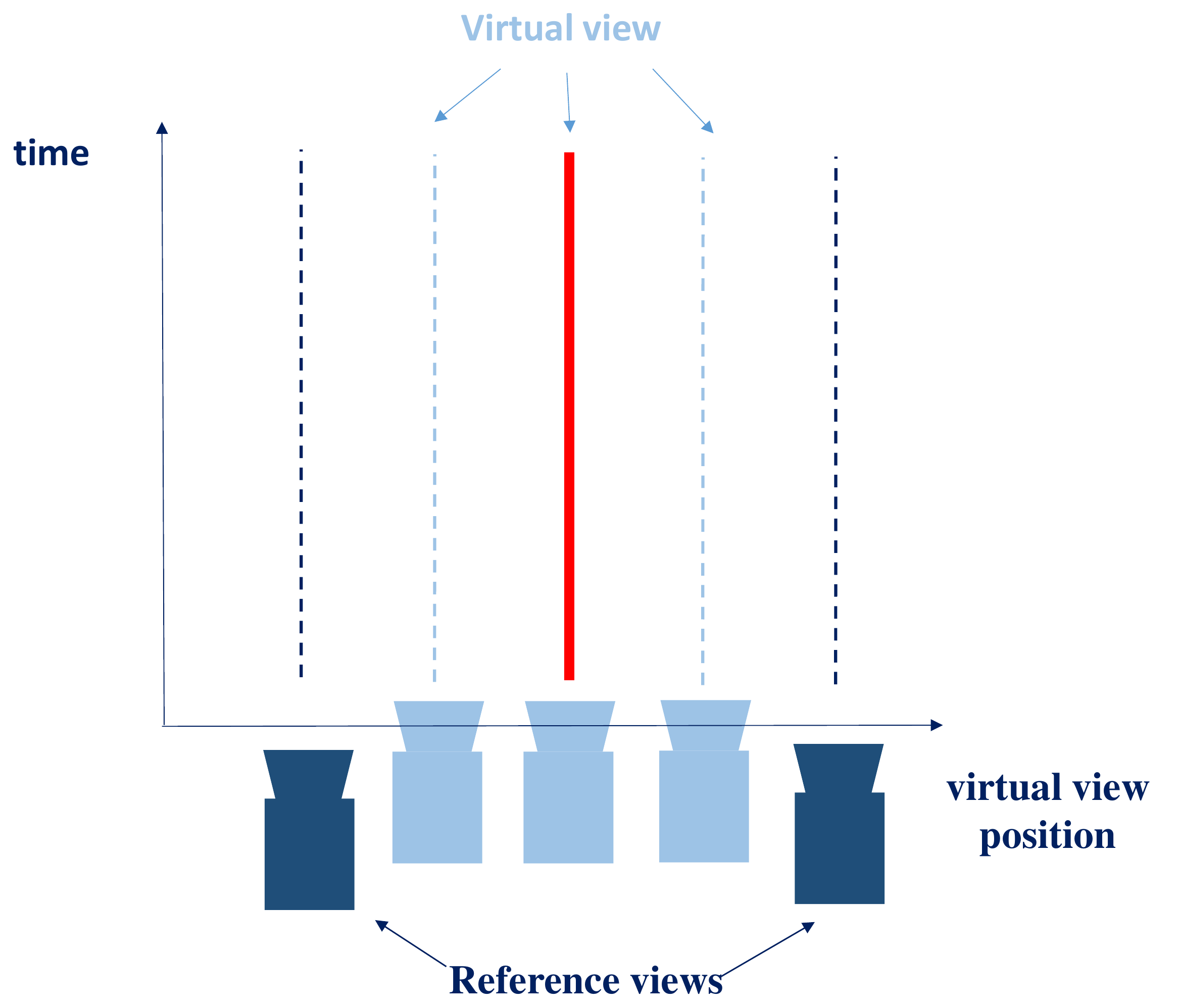}
  \end{minipage}
  \begin{minipage}[b]{0.52\linewidth}
    \centering
    \includegraphics[width=\linewidth]{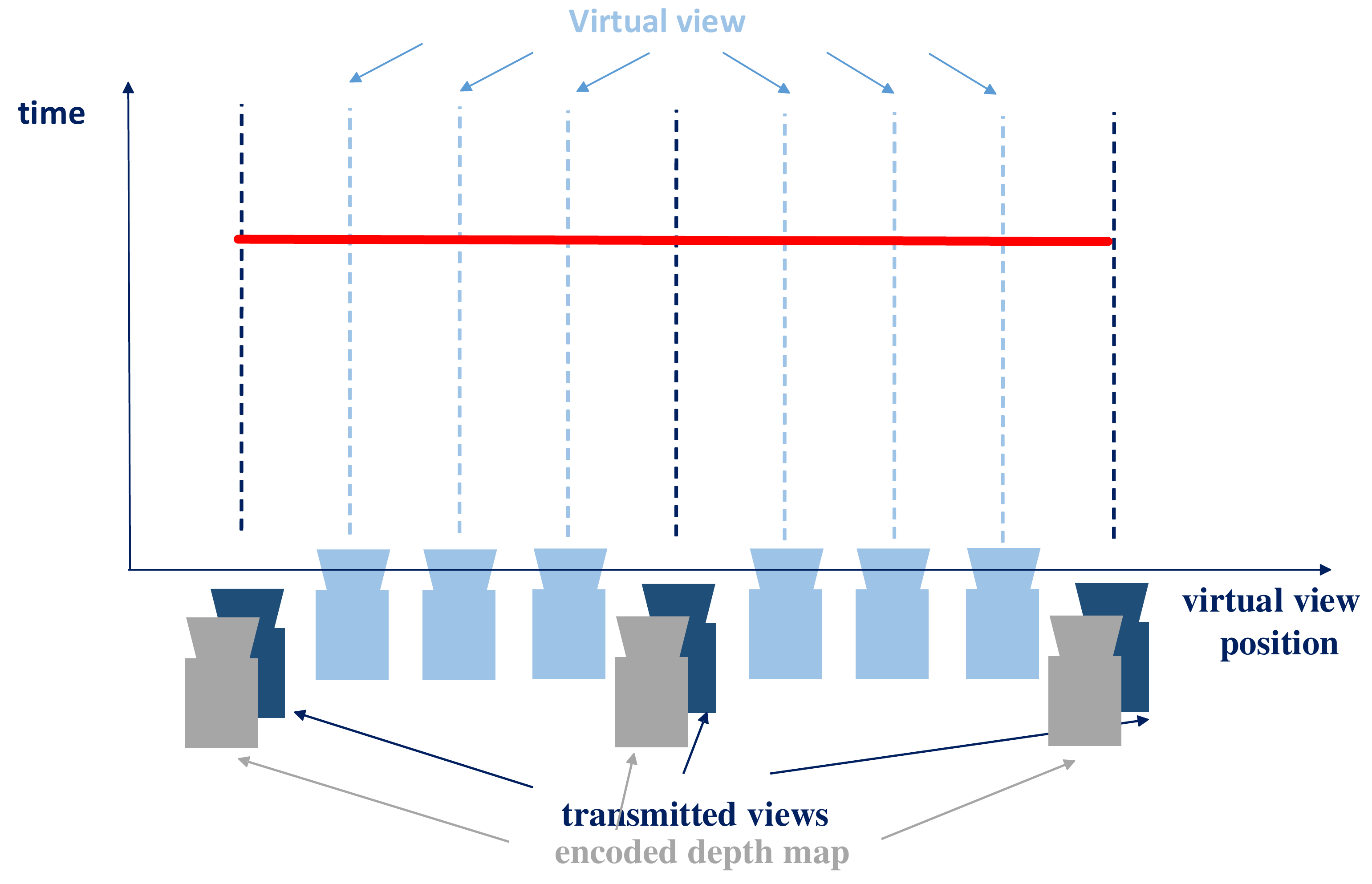}
  \end{minipage}
  \caption{Explanation of sequences' scan paths within the two datasets, where the red curves represent the virtual navigation scan-paths, the dark blue camera icons represent the reference views, the light blue camera icons represent the synthesized virtual views, and the gray camera icons represent the depth maps of the corresponding views. (a) IVC database: sequences contain temporal structure inconsistencies at one viewpoint position. (b) FVV database: sequences contain unsmooth structure transition among different viewpoint positions. }
    \label{Fig:data_scanpath}
\end{figure*}Multi-scale motion trajectory representation as spatialtemporal
distortion regions selection:

\subsubsection{Temporal structure dissimilarity} 
After getting the trajectory representations along with the extracted features, trajectories at each scale in the synthesized and reference sequences are first matched according to the averaged horizontal and vertical coordinates of the trajectories. Only the matched trajectory pairs $(t^s_{ori}, t^s_{syn})$ in the matched trajectory set $T_{m}$ would be maintained for latter deformation quantification and structure loss computation. To quantify temporal degradation by considering the two typical temporal distortions mentioned in Section~\ref{sec:Intro}, two main aspects are taken into consideration. 

First, since temporal evolution of spatial local structure-related distortions might result in deformation of motion trajectories within the sequences,~\textit{e.g.}, the motion trajectory distributed along boundaries of foreground objects might fluctuate and result in changes of the shape of the trajectory. These changes of trajectories in term of global motion trajectory deformations could be quantified by using   elastic metric described in Section~\ref{ssec:CC based on EM in ES}. More specifically, the entire deformable changes of trajectories between the synthesized and their reference sequences on all the scales $T_{EM}(T_{m})$ is defined by accumulating all the elastic errors between the trajectories: 
\begin{equation}
\label{eq:Tem}
	T_{EM}^s(T_{m})= \frac{\sum^{N_t^s}_{(t^s_{ori}, t^s_{syn}) \in T_m} D_{EM}(t^s_{ori}, t^s_{syn})}{N_t^s},
\end{equation}
 where $N_t^s$ is the number of matched trajectory pairs $(t^s_{ori}, t^s_{syn})\in T_{m}$ at scale $s$. Since $T_{EM}(\cdot)$ compute the amount of deformations between trajectories, ideally, it is able to capture not only the temporal structure-related distortions within viewpoints (at one viewpoint position) but also the one among the viewpoints (smoothness of the transition among viewpoints). 
 
As it has been pointed out in previous sections, structure-related distortions along the motion trajectories are the most disturbing temporal degradation, which could cause inconsistent transition of frames within and among viewpoints. Therefore, similar to the spatial elastic pooling stage (Section \ref{ssec:SP}), here the temporal deformation errors between each pair of matched trajectories are simply summed up to get the temporal deformation score $T_{EM}$ with Eq.( \ref{eq:Tem}). By doing so, temporal structure-related severe deformation could be well captured, while the global uniform distortions would not be over-penalized.

Second, to further quantify the non-contentiousness of transition of structure from one frame to another, structural statistical dissimilarities along trajectories are computed with the four extracted motion descriptors. More specifically, the temporal structural statistical loss $T_{SL}$ is defined as the structural dissimilarity calculated based on computing distance between matched extracted features vectors set $(H^{i,s}_{ref}, H^{i,s}_{syn}) \in H_m$ from the matched trajectory set $T_{m}$ at scale $s$: 
\begin{equation}
    T^{ij}_{SL}( H_m )=  \frac{\sum^{N_t^s}_{(H^{i,s}_{ref}, H^{i,s}_{syn}) \in H_m} D_{j}(H^{i,s}_{ref}, H^{i,s}_{syn})}{N_t^s},
\end{equation}
 where $H^{1,s}=H^s_{HOG}$, $H^{2,s}=H^s_{HOF}$, $H^{3,s}=H^s_{MBHx}$ and  $H^{4,s}=H^s_{MBHy}$ indicates the four motion descriptors and $D_j$ denotes one type of distance measures. In this paper, mainly four distance measures including  $D_1$ using Jensen-Shannon divergence (JSD), $D_2=$ Euclidean distance, $D_3=$ Cosine distance and $D_4=$ Minkowski Summation, are considered for calculating the temporal structural dissimilarity along matched trajectory base on the corresponding feature $(H^{i,s}_{syn},H^{i,s}_{ori})$.
 
\subsection{Spatial-Temporal Scores Aggregation} 
\label{sec:STA}
Finally, in order to predict the final objective score, the Support Vector Machine Regression  (SVR) is utilized to aggregate the calculated spatial elastic error $EM_{spa}$, temporal elastic error $T_{EM}$ and the 16 temporal structural errors $T^{ij}_{SL},i,j=1,...,4$ at all scales with a linear kernel. As totally seven scales are considered in this paper, the final dimension of vector representing each free-viewpoint video is 120, \textit{i.e.}, 7 scales $\times$ (16 dimension for $T_{SL}$ + 1 dimension for $T_{EM}$ ) + 1 dimension for $EM_{spa}$.  The SVR model training process is done according to ~\cite{gastaldo2013supporting,narwaria2018toward,16semantic} by employing a 1000-fold cross-validation. More specifically, for each dataset, it is randomly divided into 80\% of the videos for training and 20\% for testing, without overlap between them. 

\section{Experimental Result}
\label{sec:exp}
\subsection{Datasets}
The performance of the proposed metrics are evaluated on two datasets, including the IRCCyN/IVC DIBR Videos~\cite{bosc2011perceived} and the Free-Viewpoint Synthesized Video~\cite{bosc2013quality} dataset. In general, the first dataset contains synthesized sequences at a certain viewpoint, while the second dataset contains sequences that mimic a time-free navigation among different viewpoints. The virtual scan paths of the sequences in the two datasets are illustrated in Fig.~\ref{Fig:data_scanpath}. These two datasets contain two types of synthesized temporal structure related distortionsas mentioned in Section~\ref{sec:Intro},~\textit{i.e.}, 1) temporal structure inconsistencies at one viewpoint position and 2) unsmooth structure transition among different viewpoints, respectively. Therefore, they are selected together to benchmark the quality metrics designed for synthesized views. Detailed introductions of the two datasets are given below.

\textbf{IRCCyN/IVC DIBR Videos (IVC-DIBR)}~\cite{bosc2011perceived}: The IVC-DIBR database consists of 102 videos in resolution of $1024 \times 768$ generated with three multi-view plus depth contents. This database is designed for the evaluation of the reliability of DIBR algorithms by assessing the quality of the synthesized virtual views. Totally seven DIBR related algorithms, which are denoted as A1-A7 \cite{fehn2004depth,telea2004image,mori2009view,mueller2009view,ndjiki2011depth,koppel2010temporally}, are used to obtain 4 new virtual viewpoints for each content. It contains only synthesis related spatial-temporal artifacts within viewpoints, as there are no navigation among different viewpoints to mimic free navigation. Apart from the 9 reference sequences and the 84 synthesized virtual viewpoints, there are also 9 sequences that contain only traditional compression artifacts by encoding the texture (only videos, without compressing the depth maps) of the reference sequences. Since the purpose of this experiment is to verify the performance of metrics dedicated for capturing of related artifacts, these 9 sequences are excluded from the experiment to stress the capability of under-test metrics.  

\textbf{Free-Viewpoint Synthesized Video database (FVV)}~\cite{bosc2013quality}: The FVV database is composed of 264 videos sequences in resolution of $1024 \times 768$ /  $1920 \times 1080$ generated with six multi-view plus depth original sequences. The database is released to evaluate the impacts of depth map coding algorithms on the perceived quality of the synthesized views. Since depth maps are important during the DIBR based rendering process, seven codecs and three bitrates are adopted to encode the depth map for later synthesis process. These seven algorithms include (C1) 3D-HEVC~\cite{C1}, (C2) MVC~\cite{C2}, (C3) HM 6.1~\cite{C3}, (C4) JPEG2000~\cite{C4}, (C5) lossless-edge based codec~\cite{C5}, (C6) proposed in~\cite{C6} using color frames' correlations, and (C7) Z-LAR-RP~\cite{C7} using local information. After generating the synthesized viewpoints between the reference views with a certain configuration, sequences are then constructed with 100 key frames extracted from the synthesized viewpoints by navigating from one view to another from the left to the right and then turning around. As thus, sequences in this database contain only synthesis related spatial-temporal artifacts among viewpoints. Unlike~\cite{sandic2016free}, since blending is important in the process of DIBR based algorithm, the experiment is conducted on the entire database instead of excluding the one generated with blending mode.

\begin{table*}[!htbp]
\centering
\caption{\label{tab:main performance}%
Performance Comparison of the Proposed metric with state-of-the-art metrics. }
\label{tab:overall_performance}
\begin{tabular}
{| p{0.21\columnwidth} |  p{0.09\columnwidth}p{0.09\columnwidth}p{0.11\columnwidth}p{0.09\columnwidth}p{0.09\columnwidth}p{0.11\columnwidth} |  p{0.09\columnwidth}p{0.09\columnwidth}p{0.11\columnwidth}p{0.09\columnwidth}p{0.09\columnwidth}p{0.11\columnwidth}  | }\hline
     \bf{Dataset}  &  \multicolumn{6}{ c| }{IVC-DIBR}     &  \multicolumn{6}{ c| }{FVV}      \\ \hline 
  \bf{Metric}                &\bf{PCC$_{m}$} &\bf{SCC$_{m}$} &\bf{RMSE$_{m}$}   &\bf{PCC$_{a}$} &\bf{SCC$_{a}$} &\bf{RMSE$_{a}$}      &\bf{PCC$_{m}$} &\bf{SCC$_{m}$} &\bf{RMSE$_{m}$}   &\bf{PCC$_{a}$} &\bf{SCC$_{a}$} &\bf{RMSE$_{a}$}          \\ \hline
\multicolumn{13}{ |c| }{Image Quality Assessment Metrics (IQM) Designed for Synthesized Views} \\ \hline
3Dswim  &0.7188 & 0.6415 & 0.4394 & 0.6998 & 0.6224 & 0.4392 & 0.5587 & 0.5831 & 0.5289 & 0.5599 & 0.5786 & 0.5253 \\ \hline
MW-PSNR &0.5586 & 0.4736 & 0.5085 & 0.5305 & 0.4342 & 0.5141 & 0.4762 & 0.3769 & 0.5620 & 0.4628 & 0.3633 & 0.5653 \\\hline
MW-PSNR$_{r}$&0.5801 & 0.5292 & 0.5019 & 0.5435 & 0.4726 & 0.5082 & 0.4751 & 0.3811 & 0.5620 & 0.4579 & 0.3684 & 0.5667 \\\hline
MP-PSNR &0.6148 & 0.5791 & 0.4807 & 0.5938 & 0.5359 & 0.4849 & 0.4932 & 0.3912 & 0.5549 & 0.4730 & 0.3696 & 0.5577 \\\hline
MP-PSNR$_{r}$&0.5693 & 0.5329 & 0.5116 & 0.5532 & 0.4907 & 0.5101 & 0.4750 & 0.3791 & 0.5661 & 0.4606 & 0.3667 & 0.5664 \\\hline
 EM$_{spa}$ & 0.7200 & 0.6262 & 0.4409 & 0.6961 & 0.6059 & 0.4395 & 0.5589 & 0.5679 & 0.5345 & 0.5510 & 0.5635 & 0.5270 \\\hline
 \multicolumn{13}{ |c| }{Video Quality Assessment Metrics (VQM) Designed for Synthesized Views} \\ \hline
Liu-activity& 0.7595 & 0.6440 & 0.3978 & 0.7237 & 0.6190 & 0.4175 & 0.6413 & 0.6468 & 0.4891 & 0.6315 & 0.6159 & 0.4872 \\\hline
Liu-flicker & 0.5561 & 0.4796 & 0.5192 & 0.5470 & 0.4670 & 0.5207 & 0.6465 & 0.6463 & 0.4871 & 0.6340 & 0.6297 & 0.4880 \\\hline
 Liu-VQM   & 0.7316 & 0.6464 & 0.4188 & 0.6988 & 0.6308 & 0.4366 & 0.6676 & 0.6716 & 0.4843 & 0.6448 & 0.6233 & 0.4788 \\\hline
EM$_{tem}$ & 0.8201 & 0.8091 & 0.3021 & 0.7964 & 0.7836 & 0.3114 & 0.7756 & 0.7562 & 0.3868 & 0.7469 & 0.7464 & 0.4017 \\\hline
EM-VQM & \bf{0.8257} & \bf{0.8102} & \bf{0.3008} & \bf{0.8060} & \bf{0.7914} & \bf{0.3077} & \bf{0.7794} & \bf{0.7627} & \bf{0.3778} & \bf{0.7566} & \bf{0.7545} & \bf{0.3949} \\\hline

\end{tabular}
\end{table*}

\begin{table*}[!htbp]
\centering
\caption{Statistic significance results based on the 1000 times cross performance evaluation. For symbols in each entry of the table correspond to IVC-DIBR and FVV data set in order, \textit{i.e.}, IVC-DIBR\textbackslash{}FVV. The value `1' indicates the quality metric in the row outperform significantly the one in the column, while `-1' indicates the opposite case, and `0' indicates that the two quality metrics perform equivalently.  }
\label{tab:signifivance}
\begin{tabular}{|c|c|c|c|c|c|c|c|c|c|c|c|c|}\hline
\multirow{2}{*}{ } & 3DS    & MW-PS  & MW-PS & MP-PS & MP-PS & Liu-act & Liu-fli &  Liu-&  \multirow{2}{*}{EM$_{spa}$}  &  \multirow{2}{*}{ EM$_{tem}$}   & EM\\ 

 &  wIM      & NR$_{r}$ & NR$_{f}$ & NR$_{r}$ & NR$_{f}$ & ivity & cker &  final &   &   &  -VQA \\ \hline
 
3DSwIM             &-                    & 1\textbackslash{}1 & 1\textbackslash{}1 & 1\textbackslash{}1   & 1\textbackslash{}1  & -1\textbackslash{}-1 & 1\textbackslash{}-1 & 0\textbackslash{}-1  & 0\textbackslash{}0   &-1\textbackslash{}-1 & -1\textbackslash{}-1 \\\hline
MW-PSNR$_{f}$& -1\textbackslash{}-1 & -                  & 0\textbackslash{}0 & -1\textbackslash{}0  & -1\textbackslash{}0 & -1\textbackslash{}-1 & 0\textbackslash{}-1 & -1\textbackslash{}-1 & -1\textbackslash{}-1 & -1\textbackslash{}-1 & -1\textbackslash{}-1 \\ \hline
MW-PSNR$_{r}$& -1\textbackslash{}-1 & 0\textbackslash{}0 & -                  & -1\textbackslash{}-1 & 0\textbackslash{}0  & -1\textbackslash{}-1 & 0\textbackslash{}-1 & -1\textbackslash{}-1 & -1\textbackslash{}-1 & -1\textbackslash{}-1 & -1\textbackslash{}-1 \\\hline
MP-PSNR$_{f}$&-1\textbackslash{}-1 & 1\textbackslash{}0 & 1\textbackslash{}1 & -                    & 1\textbackslash{}1  & -1\textbackslash{}-1 & 1\textbackslash{}-1 & -1\textbackslash{}-1 & -1\textbackslash{}-1 & -1\textbackslash{}-1 & -1\textbackslash{}-1 \\\hline

MP-PSNR$_{r}$&-1\textbackslash{}-1 & 1\textbackslash{}0 & 0\textbackslash{}0 & -1\textbackslash{}-1 & -                   & -1\textbackslash{}-1 & 0\textbackslash{}-1 & -1\textbackslash{}-1 & -1\textbackslash{}-1 & -1\textbackslash{}-1 & -1\textbackslash{}-1 \\\hline
Liu-activity &1\textbackslash{}1   & 1\textbackslash{}1 & 1\textbackslash{}1 & 1\textbackslash{}1   & 1\textbackslash{}1  & -                    & 1\textbackslash{}0  & 1\textbackslash{}0   & 1\textbackslash{}1   & -1\textbackslash{}-1 & -1\textbackslash{}-1 \\\hline
Liu-flicker&-1\textbackslash{}1  & 0\textbackslash{}1 & 0\textbackslash{}1 & -1\textbackslash{}1  & 0\textbackslash{}1  & -1\textbackslash{}0  & -                   & -1\textbackslash{}0  & -1\textbackslash{}1  & -1\textbackslash{}-1 & -1\textbackslash{}-1 \\\hline 
Liu-VQM  &0\textbackslash{}1   & 1\textbackslash{}1 & 1\textbackslash{}1 & 1\textbackslash{}1   & 1\textbackslash{}1  & -1\textbackslash{}0  & 1\textbackslash{}0  & -                    & 0\textbackslash{}1   & -1\textbackslash{}-1 & -1\textbackslash{}-1 \\ \hline 
EM$_{spa}$&0\textbackslash{}0   & 1\textbackslash{}1 & 1\textbackslash{}1 & 1\textbackslash{}1   & 1\textbackslash{}1  & -1\textbackslash{}-1 & 1\textbackslash{}-1 & 0\textbackslash{}-1  & -                    & -1\textbackslash{}-1 & -1\textbackslash{}-1 \\ \hline 
EM$_{tem}$&1\textbackslash{}1   & 1\textbackslash{}1 & 1\textbackslash{}1 & 1\textbackslash{}1   & 1\textbackslash{}1  & 1\textbackslash{}1   & 1\textbackslash{}1  & 1\textbackslash{}1   & 1\textbackslash{}1   & -                    & -1\textbackslash{}0  \\\hline 
EM-VQM& 1\textbackslash{}1   & 1\textbackslash{}1 & 1\textbackslash{}1 & 1\textbackslash{}1   & 1\textbackslash{}1  & 1\textbackslash{}1   & 1\textbackslash{}1  & 1\textbackslash{}1   & 1\textbackslash{}1   & 1\textbackslash{}0   & -  \\\hline                   
\end{tabular}
\end{table*}


\subsection{Performance Evaluation Methodologies}
\label{sec:PEM}
It is emphasized in~\cite{sandic2015dibr,sandic2015dibrMP}, that commonly used image/video quality assessment metrics fail to quantify the synthesis related distortions. Therefore, the proposed metrics are only compared to the state of the art metrics designed for synthesized views in FTV scenarios as introduced in Section~\ref{sec:intro_QM}, including: 1) 3DswIM~\cite{battisti2015objective}, 2) MW-PSNR~\cite{sandic2015dibr}, 3) MP-PSNR~\cite{sandic2015dibrMP}, their reduced versions 4) MW-PSNR$_r$~\cite{sandic2016dibr}, 5) MP-PSNR$_r$~\cite{sandic2016dibr}, 6) the spatial-temporal activity distortion indicator (Liu-activity) proposed in~\cite{liu2015subjective}, 7)  the flicker distortion indicator (Liu-flicker) proposed in~\cite{liu2015subjective}, 8) and the video quality metric (Liu-VQM) proposed in~\cite{liu2015subjective} that combines Liu-activity and Liu-flicker. The predicted score of IQM, \textit{i.e.}, metrics 1)-5) and EM$_{spa}$, for one sequence are calculated by averaging the scores of all the frames.

Except for EM$_{tem}$ and EM-VQM (since SVR is already utilized), a non-linear logistic function recommended by~\cite{seshadrinathan2010study} is employed to map the objective scores $OBJ(i)$ predicted by the $i_{th}$ quality metric to the subjective quality scores before the performance evaluation. It is defined as 
\begin{equation}
\label{eq:map}
OBJ_{map}(i) =\frac{\beta_1}{1+e^{-\beta_2} \times (OBJ(i) - \beta_3)}.
\end{equation}

As described in Section~\ref{sec:STA}, support vector regression is employed to obtain the predicted quality scores of $EM_{tem}$ and EM-VQM, and the performances are computed throughout a 1000-fold cross-validation as recommended in~\cite{gastaldo2013supporting}. 
More specifically, the median and average Pearson linear Correlation Coefficient (PCC$_{m}$ and PCC$_{a}$), median and average Spearman rank order Correlation Coefficient (SCC$_{m}$ and SCC$_{a}$), as well as median and average Root Mean Squared Error (RMSE$_{m}$ and RMSE$_{a}$) between subjective and objective scores are reported across the 1000 runs for performance evaluation. For fair comparison, all the compared metrics are evaluated with the same 1000-fold cross validation, where PCC, SCC and RMSE are calculated with $OBJ_{map}(i)$ of the $i_{th}$ compared metric computed on 20\% test dataset for each fold.   

Apart from the frequently used performance evaluation methodologies, in order to better evaluate the performance of different metrics the methodology proposed by Krasula~\textit{et al.}~\cite{krasula2016accuracy,hanhart2016benchmark} is also used. In their model, it is assumed that the capability of an objective metric depends its capabilities of making reliable decisions about  1) when comparing two stimuli, whether they are qualitatively different and 2) if the are, which of them is of higher quality . The `Krasula' model is based on determining the classification capabilities of the objective models considering `Better or Worse' and `Different or Similar' scenarios.

More specifically, the capability of one objective metric to distinguish similar from significantly different pairs and the capability to indicate one stimulus is better/worse than another could determined by employing the receiver operating characteristic (ROC) analysis. Then, the performance of the metric can be verified with the area under the ROC curve (AUC) for both the `Better or Worse' and `Different or Similar' analysis, \textit{i.e.}, AUC$_{DS}$ and AUC$_{BW}$, the higher the AUC values the better metric in categorizing significantly different pairs from the similar ones as well as telling one stimulus is better/worse compared to another. (Readers are recommended to refer to~\cite{krasula2016accuracy,hanhart2016benchmark} for more detailed information.)

\subsection{Performance Comparison Results}
\subsubsection{Performance evaluation using commonly used evaluation methodologies}
Comprehensive performance evaluations of the proposed metrics are reported in this subsection. The 1000-fold cross validations results of the metrics on both the IVC-DIBR and FVV datasets are summarized in TABLE~\ref{tab:overall_performance}．

In general, according to the table, the proposed EM-VQM achieves the best performance on both the IVC-DIBR and the FVV datasets. It has a gain of 12.86\% in PCC$_m$ values on IVC-DIBR dataset and a gain of 16.75\% in PCC$_m$ values on IVC-DIBR dataset compared to the state-of-the-art video quality metric designed for synthesized videos, \textit{i.e.}, Liu-VQM. It could also be observed from the table that the objective scores predicted by most of the image quality metrics have poor correlations with subjective scores, specially on the FVV dataset. Synthesis related temporal distortions are difficult for image quality metrics to capture. Interestingly, 1) the performance of Liu-Flicker, which quantifies the amount of temporal flicker perform much better on the FVV dataset; 2) the overall performances of the quality models on FVV datasets are worse than the ones on the IVC-DIBR dataset. It could be indicated from these two observations that the second type of structure related temporal distortions in FTV scenarios, \textit{i.e.}, the unsmooth transition among viewpoints, are more challenging. By comparing the performance of EM$_{tem}$ and EM-VAM on the two datasets, it could be noticed that integrating EM$_{spa}$ does not improve the performance significantly. EM$_{tem}$ play the most important role in predicting the perceived quality.

\begin{figure*}[!htbp]
\centering
\subfloat[MP\textrm{-}PSNR]{
\includegraphics[width=0.21\textwidth]{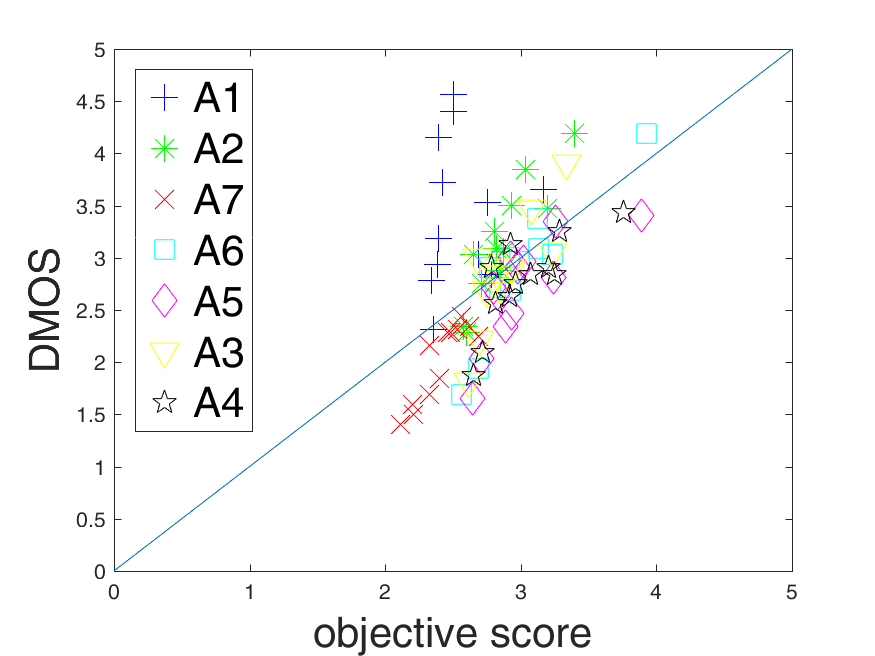}}
\subfloat[MP\textrm{-}PSNR$_{r}$]{
\includegraphics[width=0.21\textwidth]{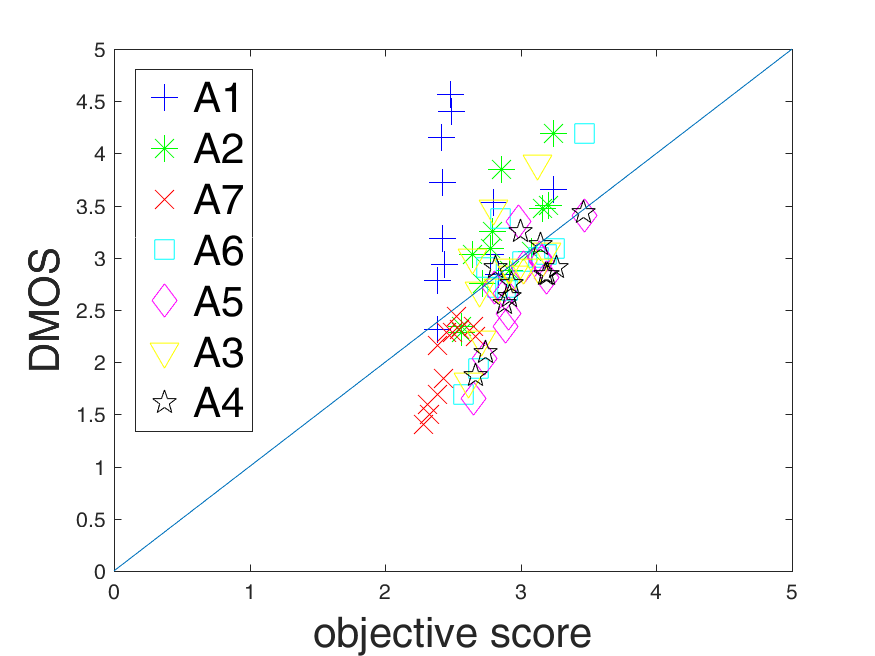}}
\subfloat[MW\textrm{-}PSNR]{
\includegraphics[width=0.21\textwidth]{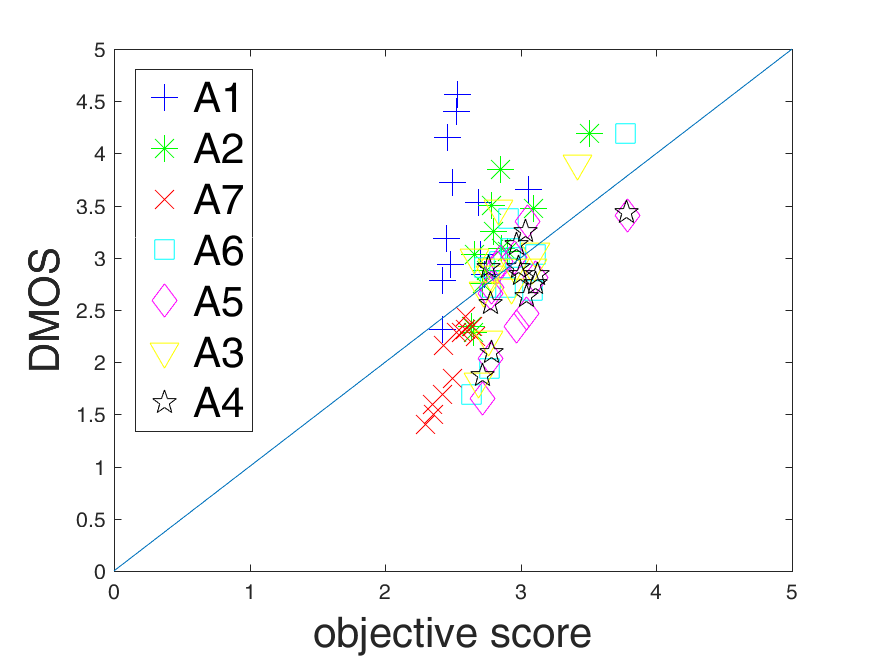}}
\subfloat[MW\textrm{-}PSNR$_{r}$]{
\includegraphics[width=0.21\textwidth]{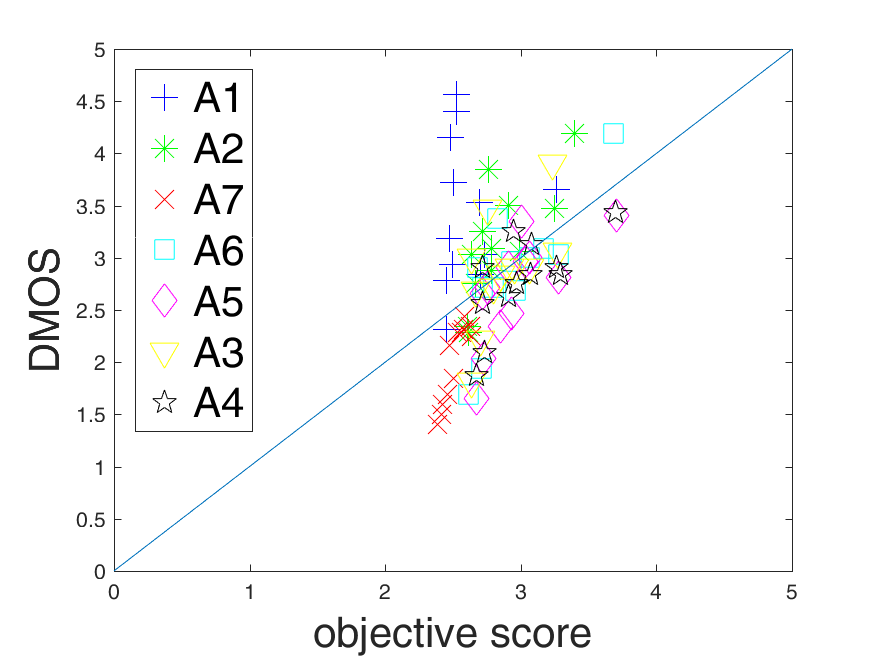}}
\\
\subfloat[3DSWIM]{
 \includegraphics[width=0.21\textwidth]{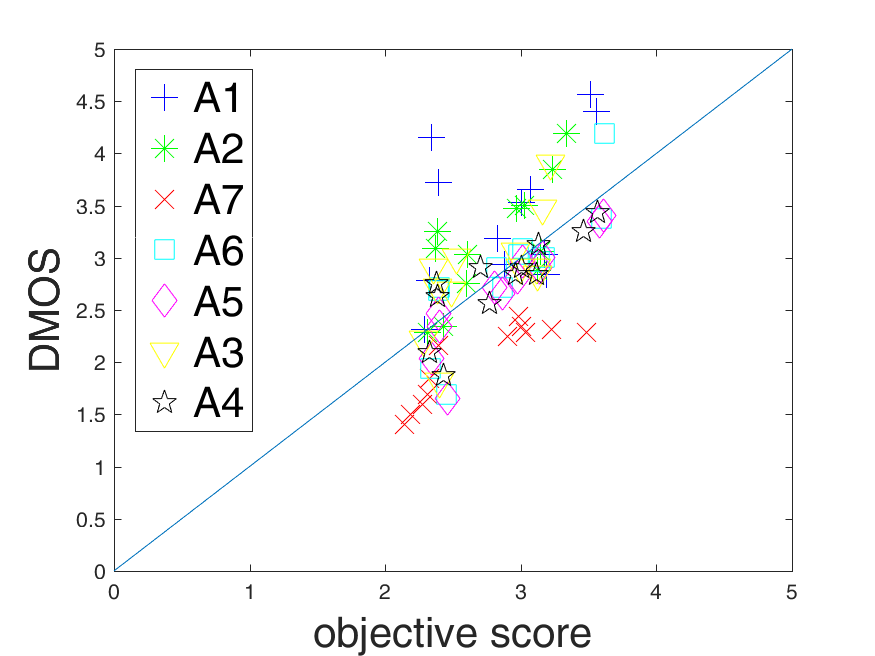}}
\subfloat[Liu-activtiy]{
\includegraphics[width=0.21\textwidth]{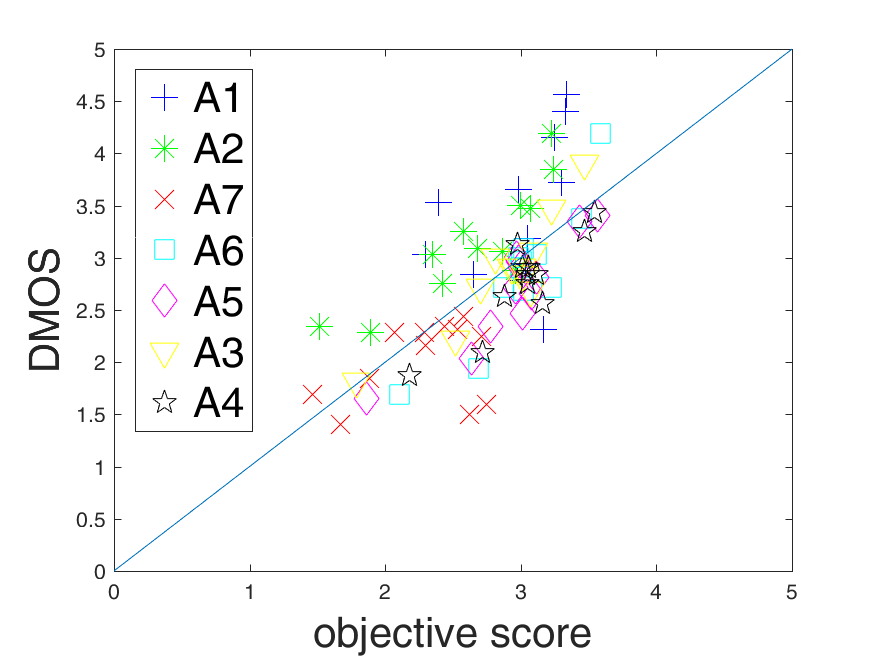}}
\subfloat[Liu-flicker]{
\includegraphics[width=0.21\textwidth]{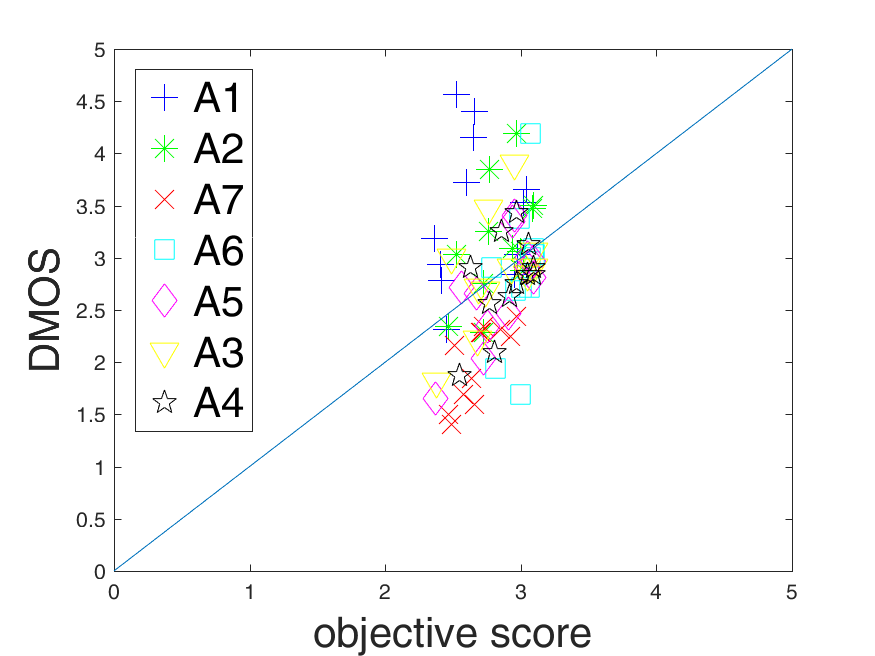}}
\subfloat[Liu-VQM]{
\includegraphics[width=0.21\textwidth]{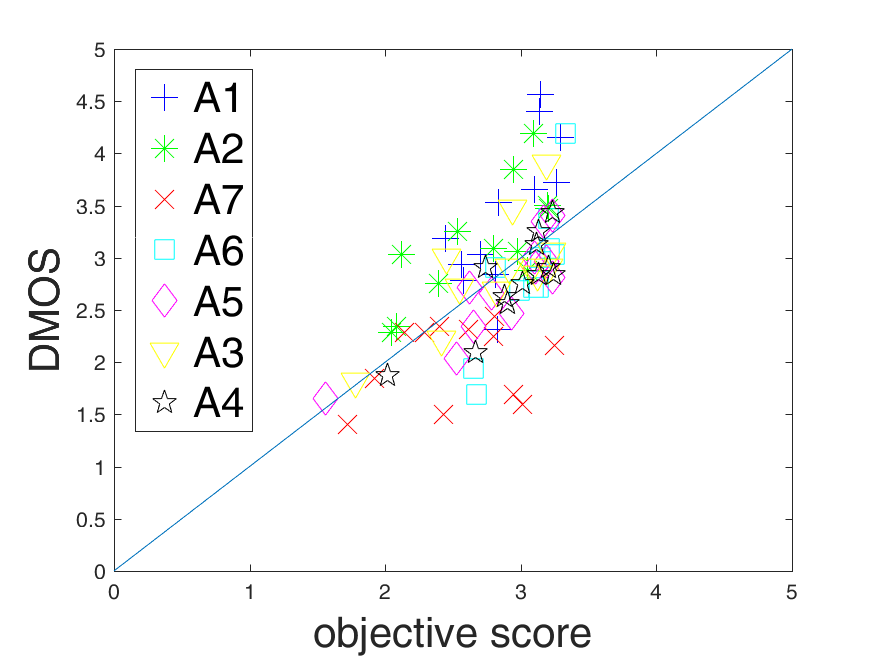}}  
\\
\subfloat[EM$_{spa}$]{
\includegraphics[width=0.21\textwidth]{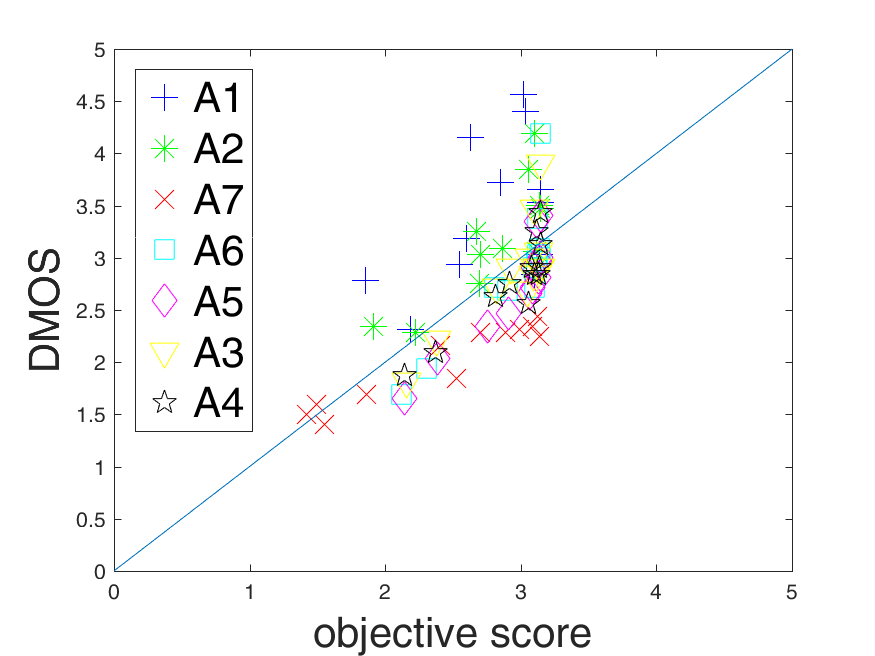}}
\subfloat[EM$_{tem}$]{
\includegraphics[width=0.21\textwidth]{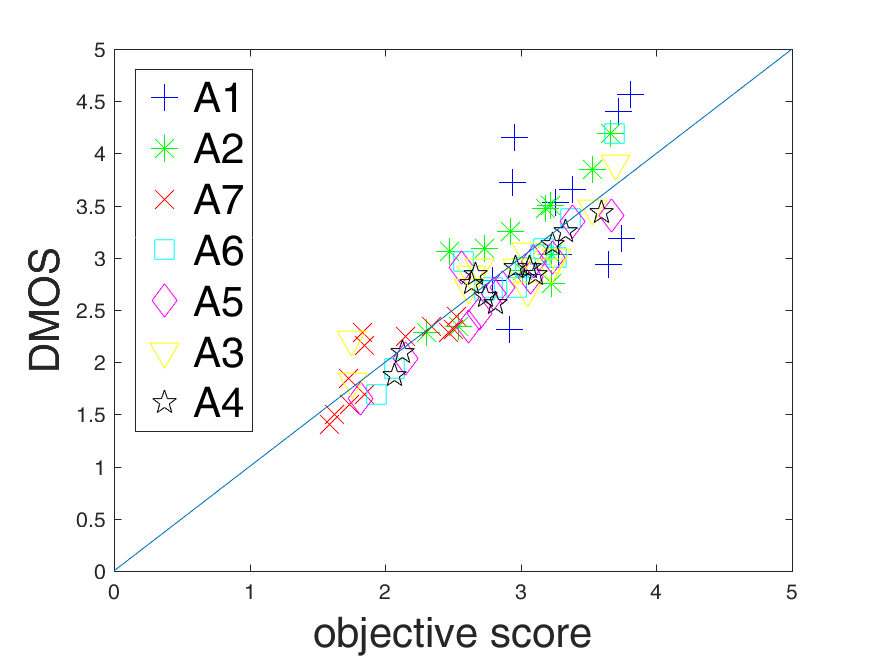}}
\subfloat[EM-VQM]{
\includegraphics[width=0.21\textwidth]{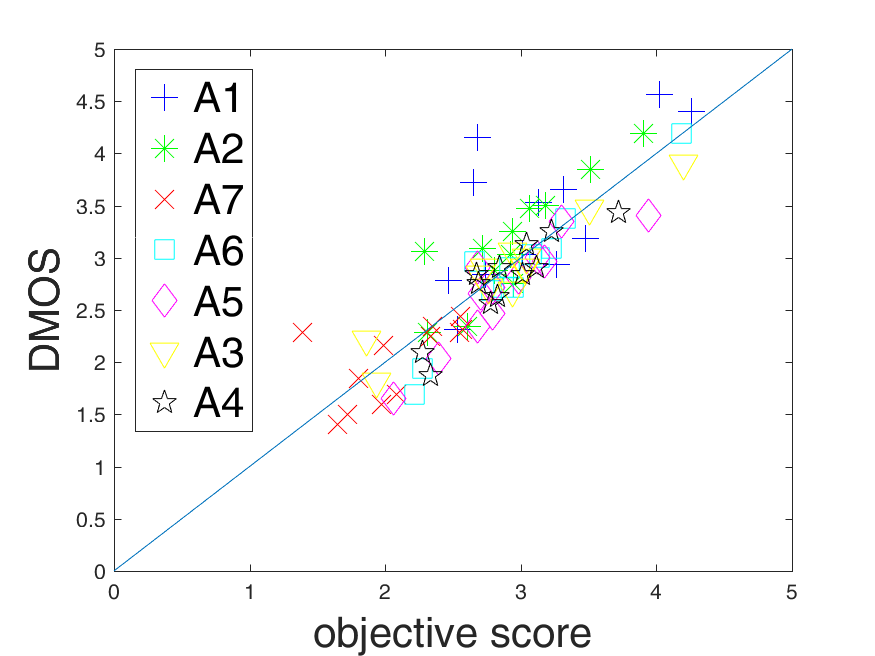}}
\caption{Scatter plots of all quality metrics' scores versus DMOS on IVC-DIBR database~\cite{bosc2011perceived}. Sequences that generated with different synthesis algorithms (\textit{i.e.}, A1-A7) are labeled with different shapes and colors (better seen in color). }
\label{fig:scatter_plot_IVC}
\end{figure*}

\begin{figure*}[!htbp] 
\centering
\subfloat[MP\textrm{-}PSNR]{
\includegraphics[width=0.21\textwidth]{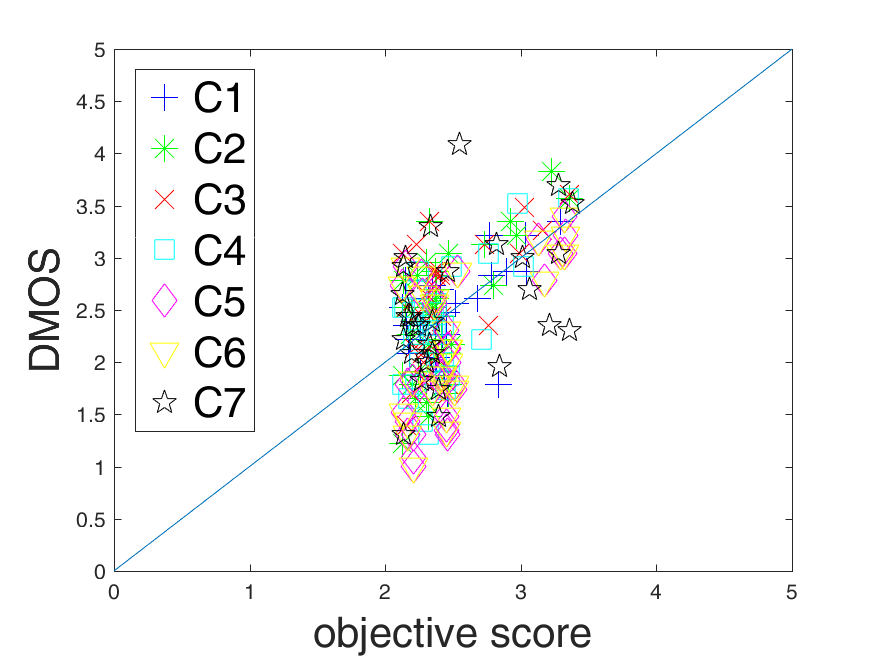}}
\subfloat[MP\textrm{-}PSNR$_{r}$]{
\includegraphics[width=0.21\textwidth]{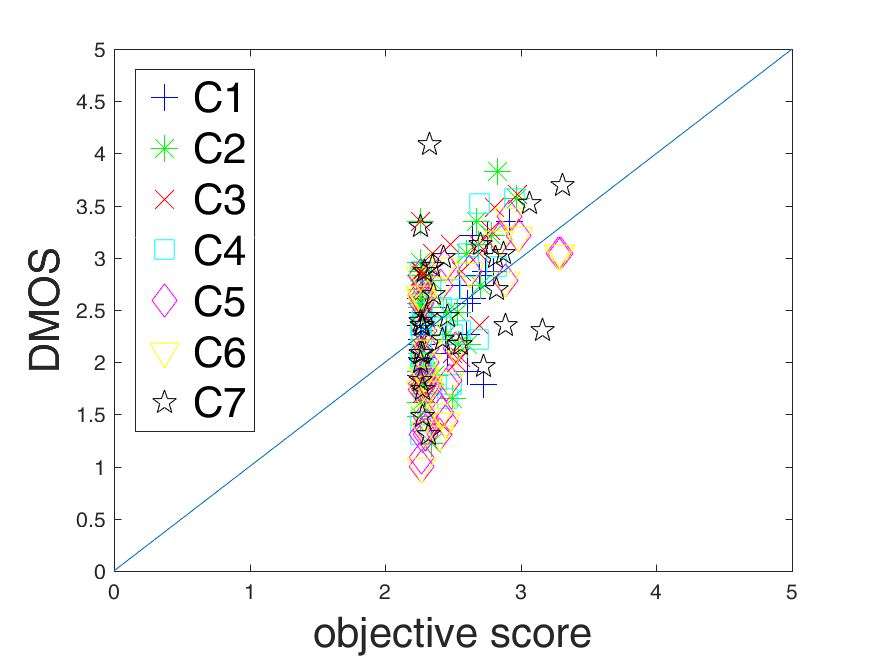}}
\subfloat[MW\textrm{-}PSNR]{
\includegraphics[width=0.21\textwidth]{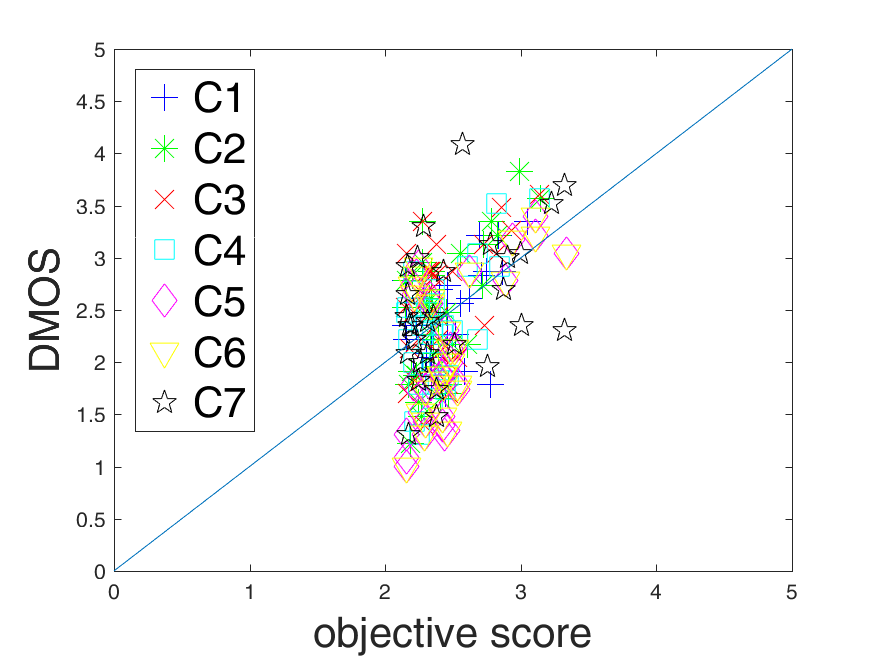}}
\subfloat[MW\textrm{-}PSNR$_{r}$]{
\includegraphics[width=0.21\textwidth]{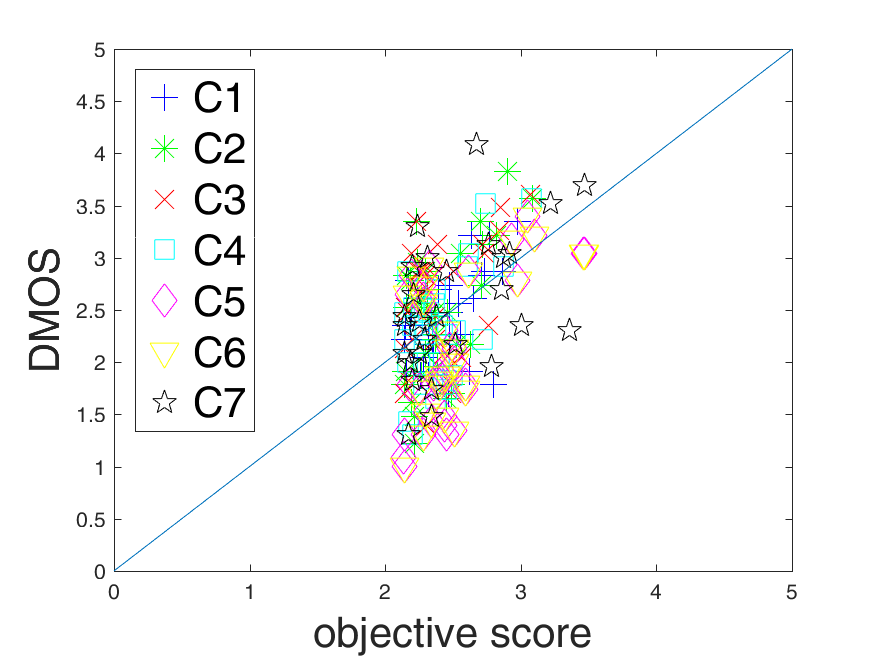}}
\\
\subfloat[3DSWIM]{
 \includegraphics[width=0.21\textwidth]{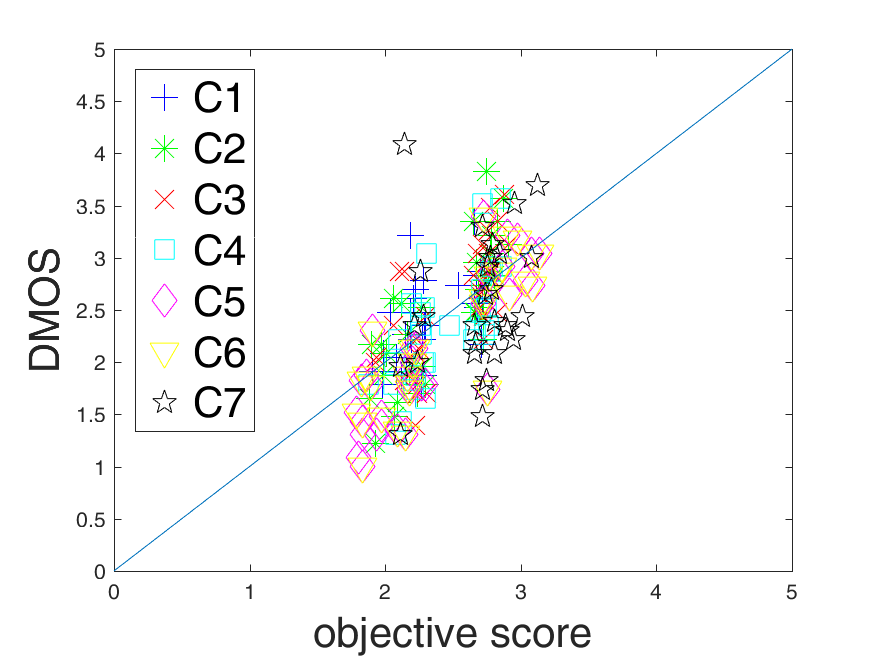}}
\subfloat[Liu-activity]{
\includegraphics[width=0.21\textwidth]{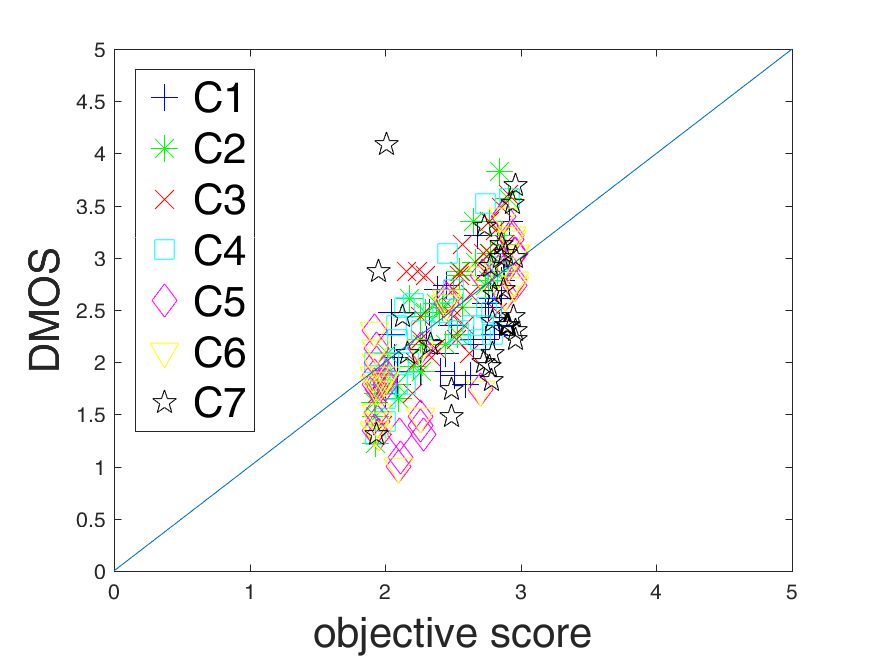}}
\subfloat[Liu-flicker]{
\includegraphics[width=0.21\textwidth]{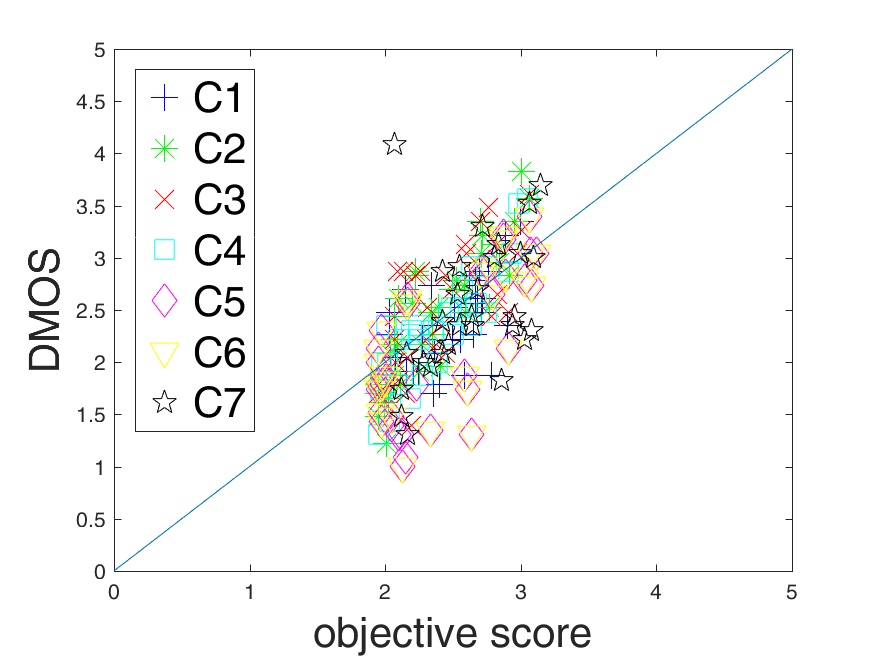}}
\subfloat[Liu-VQM]{
\includegraphics[width=0.21\textwidth]{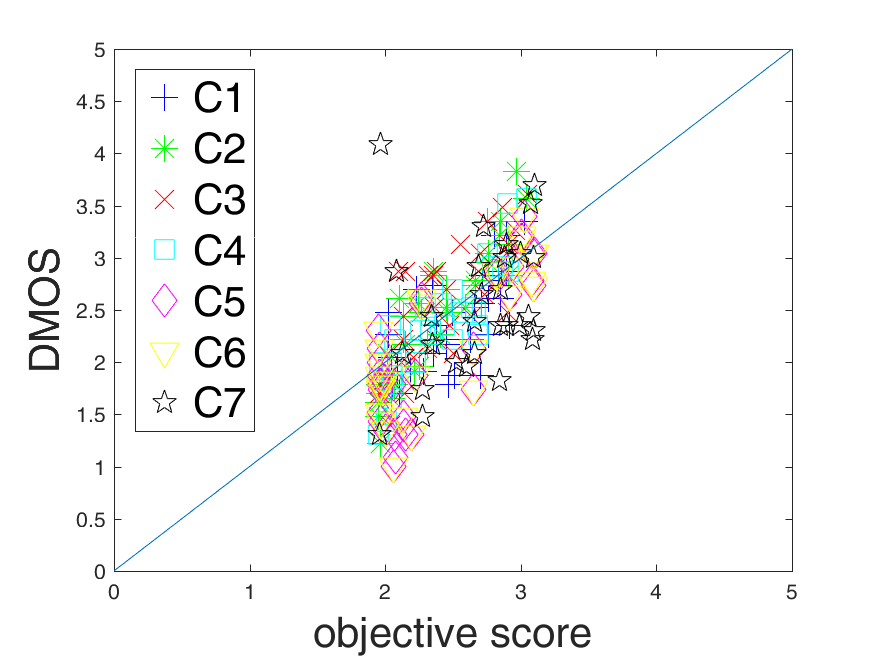}} 
\\
\centering
\subfloat[EM$_{spa}$]{
 \includegraphics[width=0.21\textwidth]{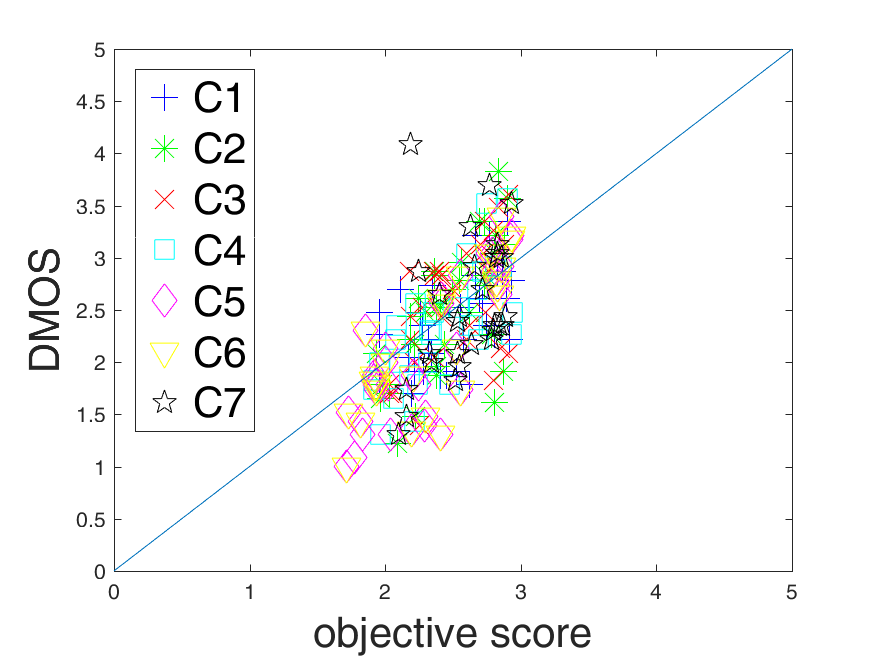}}
\subfloat[EM$_{tem}$]{
\includegraphics[width=0.21\textwidth]{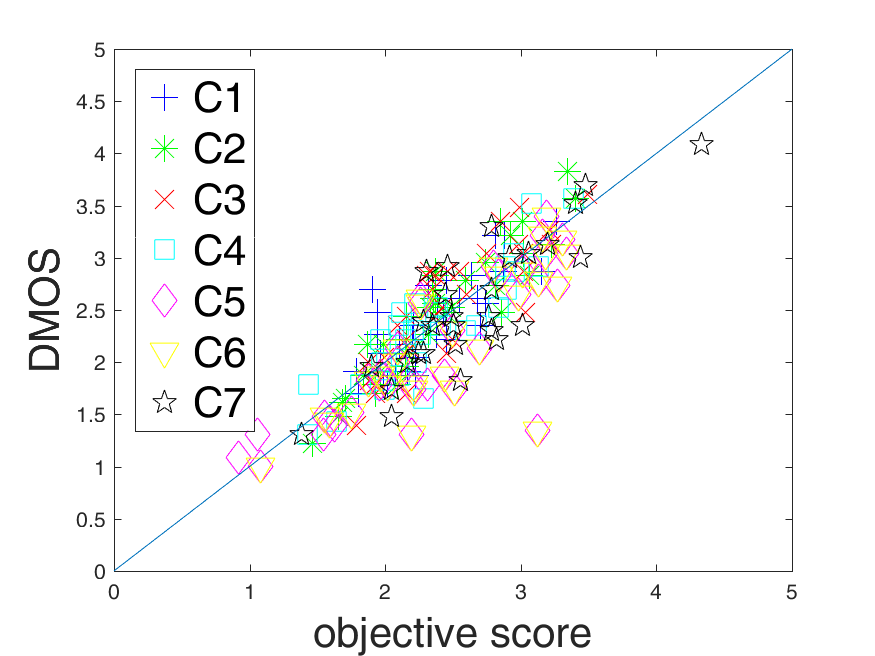}} 
\subfloat[EM-VQM]{
\includegraphics[width=0.21\textwidth]{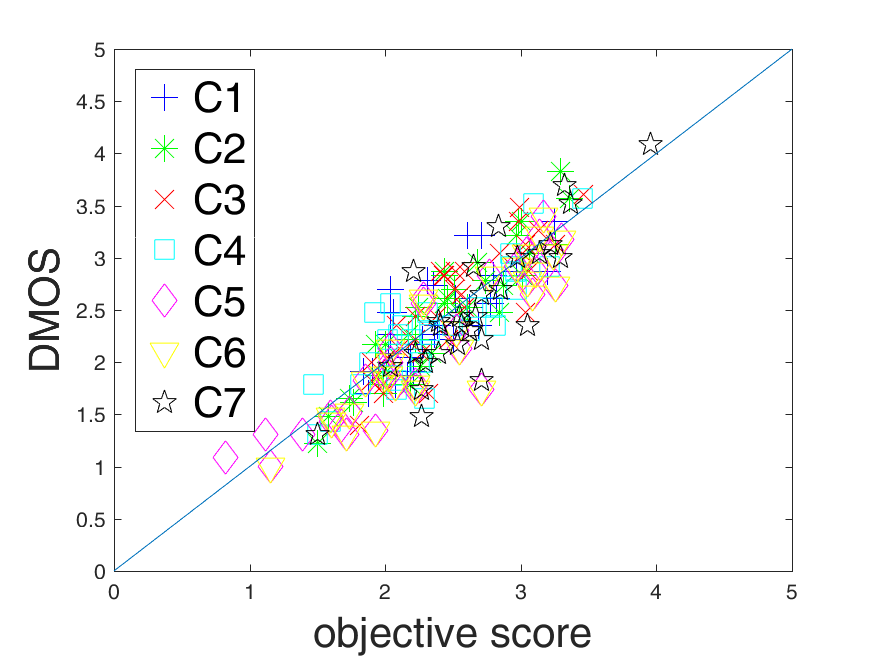}}\\

\caption{Scatter plots of all quality metrics' scores versus DMOS on FVV database~\cite{bosc2013quality}. Sequences that generated with different depth coding algorithms (\textit{i.e.}, C1-C7) are labeled with different shapes and colors (better seen in color).}
\label{fig:scatter_plot_FVV} 
\end{figure*}

The scatter plots of the tested quality metrics are illustrated in Fig.~\ref{fig:scatter_plot_IVC} and Fig.~\ref{fig:scatter_plot_FVV} respectively. For EM$_{tem}$ and EM-VQM, the model that yields the median PCC values was used to plot the figures.

In general, the quality scores predicted by both EM$_{tem}$ and EM-VQM are more consistent with DMOS compared to other image/video quality models, as most of the points shown in Fig.~\ref{fig:scatter_plot_IVC} are compactly distributed along the diagonal. It could be observed from Fig.~\ref{fig:scatter_plot_IVC} (a)-(d) that, the four point-wise PSNR based metrics tend to predict the same quality scores for sequences generated using synthesis algorithms A1. Sequences obtained using this algorithms mainly contain acceptable global shifting distortions. However, it is obvious that MP-PSNR, MP-PSNR$_r$, MW-PSNR, and MW-PSNR$_r$ over-penalize these distortions. From Fig.~\ref{fig:scatter_plot_IVC} (g), most of the sequences in IVC-DIBR dataset are predicted with scores in a range of $[2,3]$. That is because the magnitudes of flicker distortions within sequences are similar, the Liu-flicker metric could not well quantify the temporal structure inconsistencies caused by synthesis algorithms. This also explains why the performance of Liu-VQM does not outperform Liu-activity as shown in TABLE~\ref{tab:overall_performance} (since Liu-VQM is composed of Liu-activity and Liu-flicker).

According to Fig.~\ref{fig:scatter_plot_FVV}, the objective scores predicted by EM$_tem$ and EM-VQM are better aligned with the DMOS on the FVV dataset compared to the other metrics. By observing Fig.~\ref{fig:scatter_plot_FVV} (a)-(i), it could be noticed that the points are gathered as a cluster in an objective score range of $[2,3]$, and they are poorly in predicting sequences with high/bad quality. 

\subsubsection{Statistical significant test}
To examine the significance of the performances between each two tested quality metrics, student's t-test is conducted. More specifically, the 1000-fold PCC values obtained during the cross performance evaluation described above for each tested metric are used as input for t-test. The results are concluded in Table \ref{tab:signifivance} with a significance level of 0.05, where `1' represents that the performance of the under-test metric in row outperforms the one in column significantly, `-1' represents the opposite situation and `0' represents that there is no significant difference. According to the table, both the proposed EM$_{tem}$ and EM-VQM significantly outperform all the other metrics on the two datasets. In addition, EM-VQM is significantly superior to EM$_{tem}$ on IVC-DIBR dataset, but not on the FVV dataset. It reveals the fact that EM$_{spa}$ plays an important role in predicting the perceived quality score on IVC-DIBR dataset but not on the FVV dataset. Temporal distortions among viewpoints are more challenging for existing metrics, and model that considers temporal structure due to views' switch, \textit{e.g.}, EM$_{tem}$, should be considered.

\subsubsection{Performance evaluation using Krasula methodology}
The performance results of the metrics using the evaluation methodologies proposed by Krasula \textit{et al.} are reported in TABLE~\ref{tab:lukas_performance}. For EM$_{tem}$ and EM-VQM, the SVR model that obtains the median PCC values during the 1000-folds cross validation is utilized for the calculation of AUC$_{DS}$ and AUC$_{BW}$ as introduced in Section~\ref{sec:PEM}. It could be observed from the table that the proposed EM-VQM obtains the best performances, in terms of AUC$_{DS}$ and AUC$_{BW}$ values among the compared metrics. Among the compared metrics, the EM-VQM is the best in 1) distinguishing pairs of stimuli that are of similar/significantly different quality; 2) indicating sequences are of better/worse quality than others.

\begin{table}[ ]
\centering
\caption{\label{tab:lukas_performance}%
Performance Comparison of the Proposed metric with state-of-the-art metrics using the KRASULA model}
\label{tab:obj_performance}
\begin{tabular}{| c |c c  |c c|} \hline
  \bf{Dataset}  &  \multicolumn{2}{ c| }{IVC-DIBR}     &  \multicolumn{2}{ c| }{FVV}      \\ \hline 
  \bf{Metric}  & A$_{DS}$  & A$_{BW}$  & A$_{DS}$  & A$_{BW}$     \\ \hline
    \multicolumn{5}{ |c| }{IQM}      \\ \hline 
3Dswim  &0.548  & 0.830    & 0.541  & 0.775   \\  \hline 
MW-PSNR & 0.537 & 0.704  & 0.498 & 0.641  \\  \hline 
MW-PSNR$_{r}$&0.522 & 0.717   & 0.499 & 0.642   \\ \hline  
MP-PSNR &0.531 & 0.754  & 0.499 & 0.648  \\  \hline  
MP-PSNR$_{r}$&0.521 & 0.739   & 0.501 & 0.648  \\  \hline  
 EM$_{spa}$ &  0.494 & 0.830   & 0.497 & 0.761  \\  \hline 
  \multicolumn{5}{ |c| }{VQM}      \\ \hline 
 Liu-activity  &0.547 & 0.701   & 0.523 & 0.784  \\  \hline 
 Liu-flicker  & 0.473 & 0.677   & 0.514 & 0.793  \\  \hline 
 Liu-VQM   & 0.507 & 0.704  & 0.525 & 0.795  \\ \hline 
EM$_{tem}$ & 0.593 & 0.934  &   0.542 & 0.919    \\ \hline 
EM-VQM  & \bf{0.602} & \bf{0.946}     & \bf{0.543} & \bf{0.921}  \\ \hline   
\end{tabular}
\end{table}

\begin{table*}[!htbp]
\centering
\caption{ Ranking of the seven synthesis algorithms in IVC-DIBR dataset and the seven depth map codecs in FVV dataset ranked with respect to the DMOS values and predicted objective scores calculated with the image/video quality metrics designed for synthesized views. From left to right, the ranking of the synthesis algorithms/depth map codecs decreases.}
\label{tab:ranking}
\begin{tabular}{|c|ccccccc|ccccccc|}  \hline 
   \bf{Dataset}  &  \multicolumn{7}{ c| }{IVC-DIBR}     &  \multicolumn{7}{ c| }{FVV}      \\ \hline 
DMOS            & A1 & A2 & A3 & A6 & A4 & A5 & A7  & C7 & C6 & C3 & C2 & C1 & C4 & C5 \\ \hline 
3DSWIM         & A6 & A5 & A1 & A4 & A2 & A3 & A7  & C7 & C3 & C2 & C6 & C4 & C5 & C1 \\ \hline 
MW-PSNR    & A4 & A5 & A6 & A3 & A2 & A1 & A7  & C6 & C7 & C5 & C1 & C3 & C2 & C4 \\ \hline 
MW-PSNR$_r$ & A4 & A5 & A6 & A3 & A2 & A1 & A7  & C6 & C7 & C5 & C1 & C3 & C2 & C4 \\ \hline 
MP-PSNR    & A4 & A5 & A6 & A3 & A2 & A1 & A7  & C6 & C7 & C5 & C1 & C3 & C2 & C4 \\ \hline 
MP-PSNR$_r$ & A4 & A5 & A6 & A3 & A2 & A1 & A7  & C7 & C6 & C5 & C1 & C3 & C2 & C4 \\ \hline 
Liu-activity   & A4 & A6 & A1 & A5 & A3 & A2 & A7  & C7 & C1 & C3 & C2 & C6 & C4 & C5 \\ \hline 
Liu-flicker    & A6 & A4 & A5 & A2 & A3 & A7 & A1  & C7 & C1 & C6 & C3 & C5 & C2 & C4 \\ \hline 
Liu-VQM            & A6 & A4 & A1 & A5 & A3 & A2 & A7  & C7 & C1 & C3 & C6 & C2 & C5 & C4 \\ \hline 
EM$_{spa}$          & A4 & A3 & A6 & A5 & A2 & A1 & A7  & C7 & C3 & C1 & C2 & C4 & C6 & C5 \\ \hline  
EM$_{tem}$          & A1 & A2 & A6 & A4 & A3 & A5 & A7  & C7 & C6 & C3 & C2 & C1 & C5 & C4 \\ \hline 
EM-VQM          & A1 & A6 & A2 & A3 & A5 & A4 & A7  & C7 & C3 & C2 & C6 & C1 & C4 & C5 \\ \hline 
\end{tabular}
\end{table*}

\subsubsection{Benchmarking synthesis algorithms and depth map codecs}
As one of the most important functionalities of an image/video quality metric is to benchmark the performance of the system considering relative techniques. In FTV system, depth maps codecs and synthesis algorithms are two of the most important techniques. More reliable synthesis algorithms and codecs could provide better free-viewpoint videos.  From this point of view, the performances of using the objective quality metrics for benchmarking the synthesis algorithms, \textit{i.e.}, A1-A7, in IVC-DIBR dataset and the seven depth map codecs, \textit{i.e.}, C1-C7, in FVV dataset are also evaluated. More specifically, the DMOS values of sequences obtained using each synthesis algorithm/depth map codes are averaged to compute the ground truth ranking. Similarly, the predicted ranking of A1-A7/C1-C7 using the image/video quality assessment metrics are also obtained based on the mean predicted scores of each synthesis algorithm/depth map codec. The results are shown in TABLE~\ref{tab:ranking}. Comparing the ranking of A1-A7 predicted by the quality metrics with the one predicted by DMOS, the proposed EM$_{tem}$ and EM-VQM achieve the most consistent rankings. For EM-VQM, only A6 is shifted two positions forward and the positions of A4, A5 are switched. For EM$_{tem}$, only A6 is shifted one position forward and the positions of A4, A5 are switched. For most of the image quality metrics, the poor performing algorithms are ranked with higher positions, while the better performing ones are ranked lower. For the state-of-the-art DIBR-oriented video quality metric Liu-VQM, except for A7, the rest are all inconsistent with the ground truth. Comparing the ranking of C1-C7 indicated by the objective models with the ground truth, the two proposed metrics provide again the most consistent rankings. For EM$_{tem}$, only the positions of C4 and C5 are switched. For EM-VQM, the position of C6 is shifted slightly forward. Liu-activity also only incorrectly switches the position of C1 and C6, but the distance between these two codecs is far according to the ground truth. All the other compared metrics fail to provide a more correct ranking with less than three inconsistent rankings.

\section{Conclusions}
\label{sec:con}

To evaluate the quality of FVV in FTV system, in this work, we present a multi-scale motion trajectory based video quality assessment metric by quantifying the elastic changes. Specifically, to quantify the two dominant temporal structure-related distortions contained in nowadays synthesized views, \textit{i.e.}, the object deformation observed at a certain viewpoint and the structure inconsistencies observed during view switches, we calculated the amount of 1) deformation changes of spatial structure, 2) deformation changes of motion trajectories, and 3) the statistical change of motion-structure descriptors along the trajectories within reference and the synthesized videos. Then they are aggregated by SVR to predict the perceived quality. Experiments have been conducted on two databases which contain the two aforementioned temporal structure-related distortions. The results show that the proposed EM-VQM is superior to the state-of-the-art video quality metrics designed for FVV.


\balance
\bibliographystyle{IEEEtran}
\bibliography{bare_jrnl.bbl}

\ifCLASSOPTIONcaptionsoff
  \newpage
\fi

%


%
%




\end{document}